\documentclass{aa}  
\usepackage{graphicx}
\usepackage{txfonts}
\usepackage{hyperref}
\usepackage{xcolor}
\usepackage{multirow}
\usepackage{arydshln}
\usepackage[normalem]{ulem}

\newcommand\redout{\bgroup\markoverwith
{\textcolor{red}{\rule[.5ex]{2pt}{3pt}}}\ULon}

\makeatletter
\renewcommand*\aa@pageof{, page \thepage{} of \pageref*{LastPage}}
\makeatother

\begin{document}

   \title{The \textit{Gaia}-ESO Survey: Old super-metal-rich visitors from the inner Galaxy\thanks{Table XX is only available in electronic form at the CDS via anonymous ftp to cdsarc.u-strasbg.fr (130.79.128.5) or via http://cdsweb.u-strasbg.fr/cgi-bin/qcat?J/A+A/.}}


   \author{M.~L.~L.~Dantas\inst{1}
          \and
          R.~Smiljanic\inst{1}      
          \and
          R.~Boesso\inst{2, 3}
          \and
          H.~J.~Rocha-Pinto\inst{3}
          \and
          L.~Magrini\inst{4}
          \and
          G.~Guiglion\inst{5}
          \and
          G.~Tautvai{\v s}ien{\. e}\inst{6}
          \and
          G.~Gilmore\inst{7}
          \and
          S.~Randich\inst{4}
          \and
          T.~Bensby\inst{8}
          \and
          A.~Bragaglia\inst{9}
          \and
          M.~Bergemann\inst{10,11}
          \and 
          G.~Carraro\inst{12}
          \and
          P. Jofr\'e\inst{13}
          \and
          S.~Zaggia\inst{14}
          }

    \institute{Nicolaus Copernicus Astronomical Center, Polish Academy of Sciences, ul. Bartycka 18, 00-716, Warsaw, Poland\\ \email{mlldantas@protonmail.com or mlldantas@camk.edu.pl}
    \and
    Fundação Getulio Vargas’ Brazilian Institute of Economics (FGV IBRE), Rio de Janeiro, Brazil
    \and
    Observat\'orio do Valongo, Universidade Federal do Rio de Janeiro, Ladeira Pedro Ant\^onio 41, 20080-090 Rio de Janeiro, Brazil
    \and
    INAF -- Osservatorio Astrofisico di Arcetri, Largo E.Fermi, 5. 50125
    Firenze, Italy
    \and
    Leibniz-Institut f\"ur Astrophysik Potsdam (AIP), An der Sternwarte 16, 14482 Potsdam, Germany
    \and
    Institute of Theoretical Physics and Astronomy, Vilnius University, Saul\.{e}tekio av. 3, LT-10257 Vilnius, Lithuania
    \and
    Institute of Astronomy, University of Cambridge, Madingley Road, Cambridge CB3 0HA, United Kingdom
    \and
    Lund Observatory, Department of Astronomy and Theoretical Physics, Box 43, SE-221 00 Lund, Sweden
    \and 
    INAF -- Osservatorio di Astrofisica e Scienza dello Spazio di Bologna, via P. Gobetti 93/3, 40129 Bologna, Italy
    \and
    Max-Planck Institut f\"{u}r Astronomie, K\"{o}nigstuhl 17, 69117 Heidelberg, Germany
    \and 
    Niels Bohr International Academy, Niels Bohr Institute, University of Copenhagen Blegdamsvej 17, DK-2100 Copenhagen, Denmark 
    \and
    Dipartimento di Fisica e Astronomia, Universit\`a di Padova, Vicolo dell'Osservatorio 3, 35122 Padova, Italy
    \and 
    N\'ucleo Milenio ERIS \& N\'ucleo de Astronom\'ia, Facultad de Ingenier\'ia y Ciencias, Universidad Diego Portales, Ej\'ercito 441, Santiago de Chile
    \and
    INAF -- Padova Observatory, Vicolo dell'Osservatorio 5, 35122 Padova, Italy
}     

    \date{Received XXX; accepted XXX}

 
    \abstract
    {The solar vicinity is currently populated by a mix of stars with various chemo-dynamic properties, including stars with a high metallicity compared to the Sun. Dynamical processes such as churning and blurring are expected to relocate such metal-rich stars from the inner Galaxy to the solar region.}
    {We report the identification of a set of old   super-metal-rich (+0.15 $\leq$ [Fe/H] $\leq$ +0.50) dwarf stars with low eccentricity orbits ($e \lesssim 0.2$) that reach a maximum height from the Galactic plane in the range $\sim$ 0.5--1.5 kpc. We discuss their chemo-dynamic properties with the goal of understanding their potential origins.}
    {We used data from the internal Data Release 6 of the \textit{Gaia}-ESO Survey. We selected stars observed at high resolution with abundances of 21 species of 18 individual elements (i.e. 21 dimensions). We applied a hierarchical clustering algorithm to group the stars with similar chemical abundances within the complete chemical abundance space. Orbits were integrated using astrometric data from \textit{Gaia} and radial velocities from \textit{Gaia}-ESO. Stellar ages were estimated using isochrones and a Bayesian method.}
    {This set of super-metal-rich stars can be arranged into five subgroups, according to their chemical properties. Four of these groups seem to follow a chemical enrichment flow, where nearly all abundances increase in lockstep with Fe. The fifth subgroup shows different chemical characteristics. All the subgroups have the following features: median ages of the order of 7--9 Gyr (with five outlier stars of estimated younger age), solar or subsolar [Mg/Fe] ratios, maximum height from the Galactic plane in the range 0.5--1.5 kpc, low eccentricities ($e\lesssim$ 0.2), and a detachment from the expected metallicity gradient with guiding radius (which varies between $\sim$ 6 and 9 kpc for the majority of the stars).}
    {The high metallicity of our stars is incompatible with a formation in the solar neighbourhood. Their dynamic properties agree with theoretical expectations that these stars travelled from the inner Galaxy due to blurring and, more importantly, to churning. We therefore suggest that most of the stars in this population originated in the inner regions of the Milky Way (inner disc and/or the bulge) and later migrated to the solar neighbourhood. The region  where the stars originated had a complex chemical enrichment history, with contributions from supernovae types Ia and II, and possibly asymptotic giant branch stars as well.}


    \maketitle
%
\section{Introduction} \label{sec:introduction}

The stars inhabiting the solar neighbourhood provide us with  many clues on how the Milky Way (MW) develops, serving as a local laboratory that allows us to probe the Galaxy's chemo-dynamical evolution. By understanding the properties of these stellar populations, it is possible to perform a true archaeological retrieval by tracing back their assembly history \citep[e.g.][]{Bensby2003, Bergemann2014, daSilva2015, Thompson2018, Kobayashi2020}. Stellar population studies in the solar neighbourhood are very important in order to create evolutionary benchmarks not only for the MW itself, but also for other galaxies \citep[e.g.][]{VincenzoKobayashi2018}. Stellar population models built upon the stars that inhabit the MW (including the solar vicinity) are commonly used to perform stellar population synthesis with integrated light observations \citep[e.g. in stellar clusters or other galaxies; see][for reviews on this topic]{Walcher2011, Conroy2013}.

The MW is a composite of different structures, namely the thin and thick discs, the bulge, and the halo; the stars in each of these structures have particular predominant chemo-dynamic properties. The general characteristics of thin disc stars include young ages, prograde orbits of low eccentricity, low velocity dispersion, and high metallicity. The thick disc counterparts are mostly old \citep[$>$ 8 Gyr, e.g.][]{Haywood2013}, have orbits of higher eccentricity, show higher velocity dispersion, tend to be more metal-poor, and to show enhanced [$\alpha$/Fe] ratios with respect to thin disc stars  \citep[see e.g.][and references therein]{Soubiran2003, Trevisan2011, Bensby2011, Bland-Hawthorn2016}. Nevertheless, a metal-rich $\alpha$-enhanced stellar population has also been detected in the solar neighbourhood, with a possible origin in the inner disc regions \citep[e.g.][]{Adibekyan2011}. Additional controversies exist regarding stellar populations that are metal rich but posses kinematic features typical of the thick disc \citep[e.g.][]{Mishenina2004, Soubiran2005, Reddy2006, Haywood2008}. Even more striking is recent evidence that the thin disc might be older than usually assumed \citep{BeraldoeSilva2021} and that both disc components contain stars down to the extremely metal-poor regime, [Fe/H] $<$ $-$3.0 \citep{Sestito2020, DiMatteo2020, FernandezAlvar2021}.

Moreover, as a basic assumption, one might expect the chemical abundances in the interstellar medium to increase with time, as it is gradually enriched by the ejecta from successive stellar generations. It would then be possible to identify the existence of a chemical enrichment flow (CEF) of the MW \citep{Boesso2018}, a general and well-defined pace at which  chemical abundances increase with time. A combined study of abundances with stellar ages would then reveal the enrichment history of our Galaxy. However, it is now well established that there is no clear age–metallicity relation for stars near the Sun \citep[e.g.][]{Edvardsson1993, Nordstrom2004}, at least for the thin disc component (e.g. \citealt{Bensby2004} and \citealt{Haywood2006} do indicate a possible trend in the case of the thick disc). Behind such results there seems to be the fact that stars do not keep to the radii where they were formed, but might migrate significantly within the Galactic disc in their lifetimes \citep[e.g.][and references therein]{SellwoodBinney2002, Minchev2011, Roskar2013, Halle2015, Loebman2016, Khoperskov2020, Wozniak2020, Lu2022}.  

Although a unique Galactic CEF does not exist, it might still be justified to consider that separated annular regions within the disc evolve each with their own CEF. In such a flow, there might be periods when abundances increase or decrease with time \citep[such as in the case of infall of metal-poor gas; see e.g.][]{Chiappini1997, Micali2013, Spitoni2020}. If there is enough information in the form of chemical abundances and stellar ages, we can aim to separate stars formed in different radii by making reasonable assumptions regarding radial abundance gradients in the disc \citep{Minchev2018, Feltzing2020}. Within a certain sample, those stars that have been formed in regions of different CEF can potentially be identified in a multidimensional chemical space by abundance patterns that cross in some elements \citep[in the sense that one star cannot be understood as the next step of chemical enrichment following the one with a crossing pattern; see][and Section \ref{subsec:group_class} below]{Boesso2018}.

Nonetheless, stars with a given set of chemical abundances and other characteristics (such as age) may be found in orbits very different from what one would expect. One is then forced to take into account the effects of migration in an attempt to estimate the stellar birth radii \citep{Minchev2018, Feltzing2020}. Radial migration is the change of stellar orbits (i.e. guiding radius) in relation to their birthplace position, due to gravitational interaction(s), especially with non-axisymmetric structures (such as bars and transient spiral arms). The dynamical process associated with  radial migration is named ``churning'', which is characterised by a change in the angular momentum of the star, altering the orbital guiding radius without necessarily affecting the orbital eccentricity. Another dynamical process frequently discussed along with churning is ``blurring'', which is the epicyclic movement of stars in eccentric orbits from their birth radii with no change in angular momentum \citep{Schonrich2009a}. There is much debate about which of these two phenomena dominates, yet it is quite likely that stellar migration in the MW happens as a combination of both \citep[see e.g.][]{Haywood2013, Minchev2014}. Nevertheless, since churning involves a change in angular momentum, it is considered more permanent than blurring (since the star is in a more eccentric orbit and will eventually   reach its original birthplace again). Even so, according to their simulations, \citet{Halle2015} argue that the solar vicinity should not be significantly populated by migrating stars due to churning, as the Sun is just outside the outer Lindblad resonance (OLR), which acts as a barrier mitigating the migration of stars from the inner disc.

Simulations and numerical works are important to reveal features that can provide clues  to whether a star has migrated or not from the inner Galaxy. Numerous simulations show that migration is expected in MW-type galaxies, especially because of changes in the bar strength \citep[e.g.][]{BournaudCombes2002, Brunetti2011, Halle2015} and in the transient spiral arms \citep[e.g.][]{SellwoodBinney2002, Minchev2011, Roskar2012, VeraCiro2014}. The recent results of \citet{Khoperskov2020} show that the deceleration of the bar highly influences the rate of stars that escape their original birth radii. The authors also argue that metal-rich stars that migrated from the inner regions of the MW to the solar neighbourhood seem to orbit at low eccentricities, independently of whether their original eccentricities were low or large. Other dynamic features can also provide clues to whether a star has migrated or not from the inner Galaxy; for instance, \citet{DiMatteo2013} found that azimuthal variations in the metallicity distribution of old stars are a proxy for strong bar activity and, consequently, radial migration.

On the observational side, the recent large stellar spectroscopic surveys are providing increasing samples of stars that can be used to better understand radial migration. \citet{Chen2019} have found a sample of super-metal-rich stars observed by the Large Sky Area Multi-Object Fiber Spectroscopic Telescope survey \citep[LAMOST;][]{Zhao2006, Yan2022LAMOST} that seem to have migrated from the inner regions of the MW. In their findings, two  subsamples of different ages appear, one old with thick disc kinematics for which radial migration was probably important and the other  $\sim$1 Gyr in age, with thin disc kinematics and orbital features,  and likely of local origin. Studies with the Apache Point Observatory Galactic Evolution Experiment survey \citep[APOGEE;][]{Majewski2017APOGEE} have also explored the distribution of metallicity across the MW and signatures of radial migration in certain groups of stars \citep[e.g.][]{Hayden2015, Anders2017, Miglio2021, Eilers2022}. \citet{Anders2017} explored a sample of red giants and concluded that radial mixing can bring metal-rich clusters from the inner regions of the Galaxy  outwards (i.e. to Galactocentric distances of $\sim$5--8 kpc). \citet{Miglio2021} have also found a sample of old metal-rich stars, and  suggest that they were formed in the inner regions of the Galaxy and are now in the solar vicinity. It is worth mentioning that \citet{Hayden2015} argue that blurring may not be enough to explain the metallicity distribution function of the solar neighbourhood, but churning might be sufficient.

We are currently analysing the large data set of abundances for solar neighbourhood stars made available within the \textit{Gaia}-ESO Survey \citep{Gilmore2012, Randich2013, Randich2022, Gilmore2022} in an attempt to disentangle the chemical enrichment of stars formed in different regions of the disc. In the course of that analysis, we uncovered that the set of most metal-rich (MMR) stars ([Fe/H] $\gtrsim~$+0.15) have chemo-dynamic  properties that indicate a possible origin in the inner Galaxy. The purpose of this paper is to report on these stars, in particular as the \textit{Gaia}-ESO results provide one of the largest sets of abundances that has been discussed for similar stars in the literature to date \citep[see e.g.][]{Pompeia2003, Trevisan2011, Kordopatis2015, Hayden2018}.

This paper is structured as follows. In Sect. \ref{sec:sample} the sample and the methodologies are described. In Sect. \ref{sec:results} the results are discussed. Finally, in Sect. \ref{sec:conclusions}  the conclusions are summarised and the final remarks presented.

\section{Sample and methodology} \label{sec:sample}

We make use of the data products of the \textit{Gaia}-ESO internal Data Release 6 (iDR6). \textit{Gaia}-ESO is a public stellar spectroscopic survey that observed over 110,000 stars with the Fibre Large Array Multi  Element Spectrograph \citep[FLAMES;][]{Pasquini2002}, the multi-fibre facility at the Very Large Telescope in Cerro Paranal, Chile. The overall sample considered for the present analysis contains 3928 stars marked with the keyword \texttt{GES\_TYPE} equal to \texttt{GE\_MW}. These are field dwarfs within $\sim$2 kpc of the Sun (see Fig. \ref{fig:heliocentric_distances}) observed using the Ultraviolet and Visual Echelle Spectrograph \citep[UVES,][]{Dekker2000} with a resolving power R=47,000 and covering a spectral range 480.0--680.0~nm. The selection function of these observations is explained in \citet{Stonkute2016}. Dwarfs observed with UVES were selected to have 2MASS \citep{2MASS} $J$ magnitudes mostly in the range 12--14 mag. About 75\% of the final targets in our sample are hotter than 5600 K (spectral type G6V). This means that the majority of our stars are at distances larger than about 430 pc (50\% of them in the range 430--1340 pc, assuming the absence of reddening). In addition, we include a smaller sample of 315 field stars observed with UVES towards the Galactic bulge (\texttt{GES\_TYPE} keyword equal to \texttt{GE\_MW\_BL}), but that are most likely within the inner disc.\footnote{The fainter stars in these fields, observed with the GIRAFFE spectrograph, are likely bulge stars. The UVES fibres, however, were allocated to brighter objects that are not expected to be at the distance of the bulge.} Stars observed in the field of open or globular clusters are not included in our sample. These stars would naturally clump together in chemical space and affect our capability of recovering an unbiased picture of the chemical enrichment of the solar neighbourhood. The detailed description of the sample selection   follows, and the final sample is briefly described in Sect. \ref{subsec:final_sample}.

\subsection{Data selection} \label{subsec:dataselec}

The sample was restricted to stars with measured abundances of all the following species (i.e. discarding those with missing data, \texttt{NAN}): \ion{C}{i}, \ion{Na}{i}, \ion{Mg}{i}, \ion{Al}{i}, \ion{Si}{i}, \ion{Si}{ii}, \ion{Ca}{i}, \ion{Sc}{ii}, \ion{Ti}{i}, \ion{Ti}{ii}, \ion{V}{i}, \ion{Cr}{i}, \ion{Cr}{ii}, \ion{Mn}{i}, Fe,\footnote{The abundance of Fe was estimated via [Fe/H] provided by the iDR6 of the \textit{Gaia}-ESO survey. Therefore, there is no ionisation level associated with this abundance.} \ion{Co}{i}, \ion{Ni}{i}, \ion{Cu}{i}, \ion{Zn}{i}, \ion{Y}{ii}, \ion{Ba}{ii}. The choice of these abundances was made in order to maximise the number of stars with as many measurements as possible. For the solar abundance scale, we used abundances from \citet{Grevesse2007}. The abundances were computed with the codes described in \citet{Smiljanic2014} and combined with the Bayesian methodology described in Worley et al. (in prep.; see also a summary in \citealt{Gilmore2022}).

Further selection criteria include the removal of stars with signal-to-noise ratio (\texttt{S/N}) lower than 40, rotating faster than \texttt{VSINI}$>$10 km s$^{-1}$, and with any peculiar flags (e.g. binarity, emission lines, asymmetric line profiles); in other words, by using the flag \texttt{PECULI}, all stars that had any values different from \texttt{NAN} were removed. We used the internal cross-match between \textit{Gaia}-ESO sources and the \textit{Gaia} early Data Release 3 (EDR3) catalogue \citep{Gaia2016, GaiaEDR3} to extract astrometric parameters for the sample. The stars with \texttt{parallax}$\leq$0 were discarded. Finally, only those with the following flags were selected: \texttt{ruwe}$<$1.4, \texttt{ipd\_frac\_multi\_peak}$\leq$2, and \texttt{ipd\_gof\_harmonic\_amplitude}$<$0.1, as recommended by \citet[][Sect. 3.3]{Fabricius2021}.

\subsection{Orbits and ages} \label{subsec:orbits_age}

To integrate the orbits of the stars we   made use of the radial velocities from {\em Gaia}-ESO and astrometric parameters from \textit{Gaia} EDR3 (i.e. parallax, proper motions, and their respective associated uncertainties). The parallax zero-point correction was made according to the prescription of \citet{Lindegren2021}. A total of 24 stars were discarded in this step as they did not meet the requirements to calculate the zero-point correction.\footnote{For more information, please see: \url{https://gitlab.com/icc-ub/public/gaiadr3_zeropoint/-/blob/master/tutorial/ZeroPoint_examples.ipynb}}: \texttt{pseudocolour}=\texttt{NAN}; or 1.24 $\leq$ \texttt{pseudocolour} $\leq$ 1.72; and \texttt{astrometric\_params\_solved}$>$3.

Distances were estimated using the corrected parallaxes and the Bayesian recipe from \citet{Bailer-Jones2015}.\footnote{Using the \textsc{r} code available on \url{https://github.com/ehalley/parallax-tutorial-2018/}.} The next step is the integration of orbits, which was done with \textsc{galpy}, a \textsc{python} package for Galactic dynamics \citep{Bovy2015}, which also makes use of \textsc{astropy} \citep{Astropy2013, Astropy2018}. It is worth mentioning that \textsc{galpy} uses left-hand galactocentric coordinates as reference.\footnote{As informed in the documentation: \url{https://docs.galpy.org/en/v1.7.0/getting\_started.html\#orbit-integration}.} To run \textsc{galpy} we used the MW potential proposed by \citet{McMillan2017}, a 10 Gyr period, and \texttt{method=`dop853\_c'}, which stands for the Dormand-Prince integration in \textsc{C} \citep{DormandPrince1980}. We describe step-by-step how the uncertainties of \textsc{galpy}'s output parameters were estimated as follows.

In the first step each input parameter was re-sampled individually 100 times and their distribution was considered Gaussian. This was done to estimate the covariance among all parameters used as input by \textsc{galpy}, which are right ascension (\texttt{ra}), declination (\texttt{dec}), proper motions for both \texttt{ra} and \texttt{dec}, radial velocity, and median distance (the last provided by the  Bayesian distance code), and their respective uncertainties. In step 2, with the first re-sampling at hand, we estimated the covariance among these parameters. In step 3 we re-sampled the parameters 100 times considering a multivariate Gaussian with the covariance matrix determined in the previous step. Finally, in step 4 we ran \textsc{galpy} for each of multivariate re-sampled star (in other words, 100 times for each star), which allowed us to retrieve the confidence intervals for all its output parameters.

For age estimations we made use of \textsc{unidam}, a Bayesian \textsc{python} package for isochrone fitting \citep{Mints2017, Mints2018} that relies on PARSEC evolutionary tracks \citep{Bressan2012} to perform this fit. As input, \textsc{unidam} makes use of the effective temperature ($T_{\rm{eff}}$), surface gravity ($\log(g)$), [Fe/H], a combination of 2MASS $JHK_s$ \citep{2MASS} and AllWise $W1W2$ \citep{AllWISE} magnitudes, and all their respective uncertainties. In some cases \textsc{unidam} provides more than one potential solution for a given star. We carefully analysed the stars with multiple solutions and realised that \textsc{unidam} provides bimodal results for them:  one set of ages that is very young (with $\log(t)$ centred at $\sim$7) and another that is quite old ($\log(t)$ centred at $\sim$10). Further checks suggested that these solutions are degenerate and the values still quite uncertain. We finally decided to discard from the sample all the stars with more than one possible age solution. Stars with only one solution have the \texttt{quality} flag equal to 1; all other results for the \texttt{quality} flag were discarded.

\subsection{The final sample}  \label{subsec:final_sample}

The final sample is comprised of 1460 stars, all with orbits, ages, and abundances for the species listed in Section \ref{subsec:dataselec}.\footnote{The catalogue can be accessed in electronic form at the CDS.}  A total of 171 stars are in what we define as the MMR group; the  metallicity range and summary statistics are available in Table \ref{tab:mmr_subgs}. In the MMR group, 170 are in fields marked as \texttt{GE\_MW} and 1 as \texttt{GE\_MW\_BL}. In this paper we discuss only the properties of the MMR group.

\begin{table*}
    \caption{Summary statistics for the metallicities of the MMR subgroups.}
    \centering
    \begin{tabular}{l|c|c|c|c|c|c|c}
    MMR subgroup & \# of $\star$s & $\langle \rm{[Fe/H]} \rangle$ & $\sigma_{\rm{[Fe/H]}}$ & $\rm{[Fe/H]_{max}}$ & $\rm{[Fe/H]_{min}}$ & $\langle \rm{[Mg/Fe]} \rangle$ & $\sigma_{\rm{[Mg/Fe]}}$ \\
    \hline
    \hline
     7 (dark green)  & 61 & 0.27 & 0.035 & 0.35 & 0.17 &  -0.050 & 0.066 \\
     6 (orange)      & 17 & 0.29 & 0.049 & 0.33 & 0.15 &  -0.030 & 0.070 \\
     8 (pink)        & 59 & 0.34 & 0.034 & 0.40 & 0.24 &  -0.040 & 0.057 \\
     9 (light green) & 24 & 0.39 & 0.033 & 0.47 & 0.31 &  -0.005 & 0.048 \\
    10 (purple)      & 10 & 0.49 & 0.049 & 0.50 & 0.35 &  -0.010 & 0.083 
    \end{tabular}
\label{tab:mmr_subgs}
\tablefoot{We display their hierarchical clustering (HC) classification number (i.e. MMR subgroup number, which we adopt throughout this paper) and colour associated in all plots, the number of stars in each group, median and standard deviations of both [Fe/H] and [Mg/Fe], maximum and minimum values for [Fe/H]. The order displayed is with increasing $\langle \rm{[Fe/H]} \rangle$.}
\end{table*}

\begin{table}
    \caption{The three different levels of clustering that we performed in our analysis.}
    \centering
    \begin{tabular}{c|c|c|l}
         HC level & $d_W$ & \# of (sub)groups & Comments\\
         \hline
         \hline
         Level 1 & 6.0   & 6  & Main large groups \\
         Level 2 & 3.5   & 11 & Subgroups \\
         Level 3 & 1.3   & 48 & Subgroups (MMR)
    \end{tabular}
    \label{tab:hc_levels}
    \tablefoot{In this work we mainly use the main groups. Level 1 to select the MMR and Solar groups; and Level 3, which we use to select the subgroups of the MMR group.}
\end{table}

\begin{table*}
    \caption{Summary statistics for the ages of the MMR subgroups.}
    \centering
    \begin{tabular}{l|c|c|c|c|c|c|l|l}
        MMR subgroup & $\overline{\log(t)}_{50\%}$ & $\overline{t}_{50\%}$& $\overline{\log(t)}_{84\%}$ & $\overline{\log(t)}_{16\%}$ & $\overline{\log(t)}_{\rm{max}}$ & $\overline{t}_{\rm{max}}$ & $\overline{\log(t)}_{\rm{min}}$&  $\overline{t}_{\rm{min}}$\\
        & & \footnotesize{(Gyr)} & & & & \footnotesize{(Gyr)} & & \footnotesize{(Gyr)} \\
        \hline
        \hline
         7 (dark green)  & 9.84 & 6.92 &  9.91 & 9.79 & 10.11 & 12.88 &  9.37(*) &  2.34(*) \\
         6 (orange)      & 9.95 & 8.91 & 10.02 & 9.74 & 10.13 & 13.49 &  9.47(*) &  2.95(*) \\
         8 (pink)        & 9.91 & 8.13 &  9.96 & 9.86 & 10.09 & 12.30 &  9.51    &  3.24    \\
         9 (light green) & 9.89 & 7.76 &  9.96 & 9.85 &  9.99 &  9.77 &  9.55    &  3.55    \\
        10 (purple)      & 9.96 & 9.12 &  9.98 & 9.93 & 10.11 & 12.88 &  9.77    &  5.89 
    \end{tabular}
    \label{tab:mmr_subgs_ages}
    \tablefoot{We display their hierarchical clustering (HC) classification number (i.e. MMR subgroup number, adopted throughout this paper) and colour associated in all plots, and age fitting results (both for $\log(t)$ and $t$) retrieved from \textsc{unidam}. For the ages, we display the median of $\overline{\log(t)}$ and $\overline{t}$, i.e. 50\textsuperscript{th} percentile for each; 16\textsuperscript{th} and 84\textsuperscript{th} percentiles (corresponding to $\mp 1\sigma$ interval) for $\overline{\log(t)}$, and maximum and minimum values for $\overline{\log(t)}$ and $\overline{t}$, respectively. The order displayed is with increasing $\langle \rm{[Fe/H]} \rangle$. (*) Groups 6 and 7 have, respectively, 2 and 3 outliers of younger age, which we remove from the analysis (as well as from this Table); see Fig. \ref{fig:appendix_agedistribution}.}
\end{table*}

\subsection{Stellar group classification} \label{subsec:group_class}

We then proceeded to divide the stars  with similar chemical compositions into groups. The goals were twofold. One was to explore whether we can find groups of stars that are chemically homogeneous and might have originated from the same Galactic region by analysing their dynamic properties and age distribution. The other goal was to test whether such groups can be ordered in some sort of evolutionary sequence, which means that the groups would be enriched at `levels', following a logical sequence (i.e. groups that are more metal-rich  are also younger), resulting in a CEF. For that division we made use of only the chemical abundances listed in Sect. \ref{subsec:dataselec}. With these abundances at hand, we applied the hierarchical clustering \citep[HC; e.g.][]{Murtagh&Contreras2012, Murtagh2014}, by making use of \textsc{scipy} \textsc{python} package \citep{2020SciPy-NMeth}, similarly to what was done by \citet{Boesso2018}. 

\begin{figure*}
    \centering
    \includegraphics[width=\linewidth]{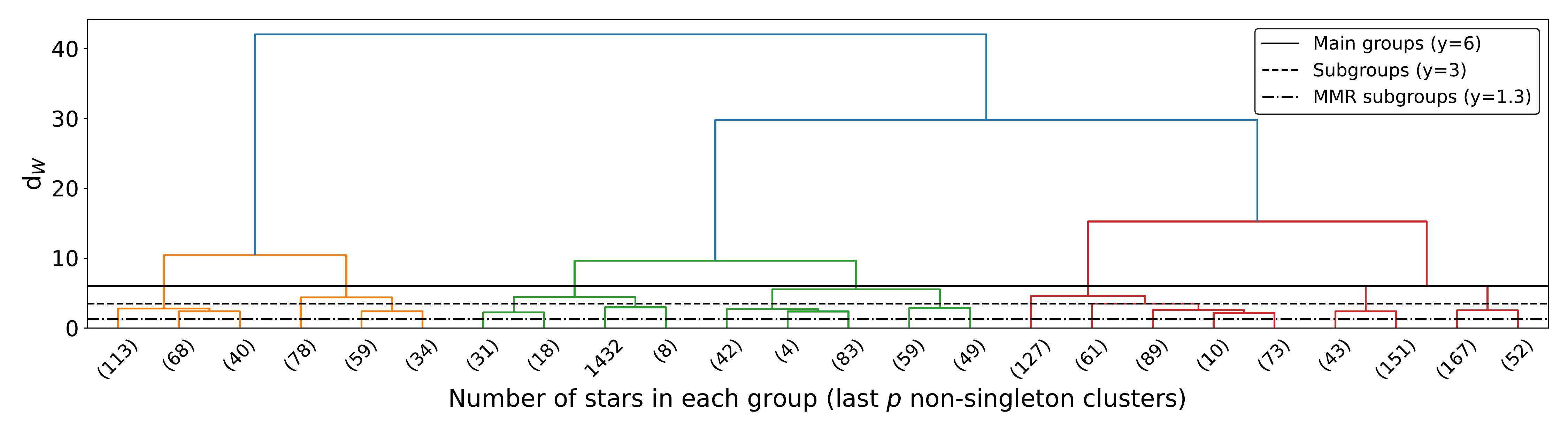}
    \includegraphics[width=\linewidth]{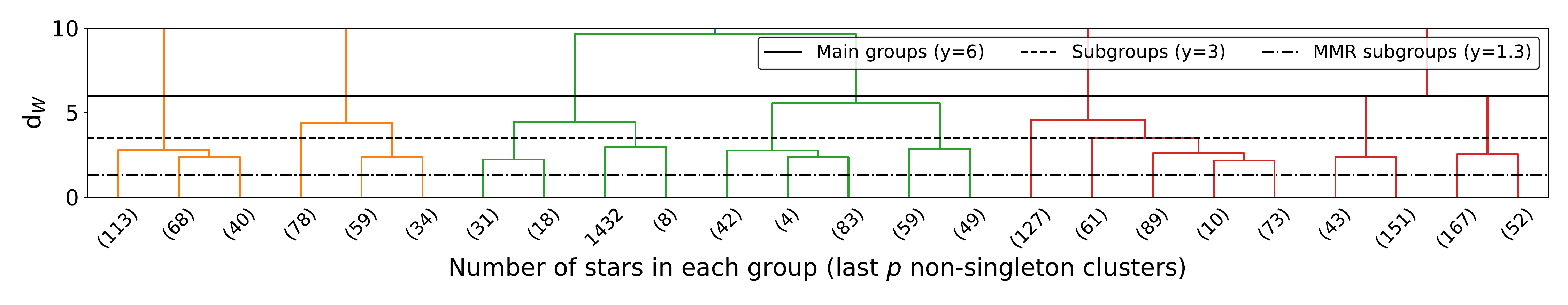}
    \caption{Dendrogram with the hierarchical classification of the stars into groups. The solid line is for the main groups with Ward distance $d_W$=6; the dashed line is for the general subgroups with $d_W$=3. The   dot-dashed line is for the  MMR subgroups,   established with $d_W$=1.3. The end of the dendrogram is truncated with the total number  of stars in each branch shown on the $x$-axis. The numbers that appear on the $x$-axis are the final groups automatically detected by the dendrogram using the last $p$ non-singleton clusters ($p=24$)  to facilitate   visualisation \citep[for more details, see the algorithm description in the \textsc{scipy} package,][]{2020SciPy-NMeth}. The number of stars in each MMR subgroup can also be seen in Table \ref{tab:mmr_subgs}. \textit{Top panel:} Full dendrogram. \textit{Bottom panel:} Zoomed-in image of dendrogram  to display the  range $0 \leq d_{W} \leq 10$.}
    \label{fig:dendrogram}
\end{figure*}

Hierarchical clustering is a non-parametric method for groupings in multivariate data, which allows us to assemble and label stars according to their abundances without any previous assumptions. Nonetheless, it is possible to set some customisation in the HC, such as the type of distance used in the clustering algorithm. In this case we chose the Ward distance (hereafter $d_W$), also known as Ward's minimum variance method \citep{Ward1963}. In this HC method, the choice of clusters to be merged is made in such a way that the intra-cluster summed squared distance is minimised. Euclidean distance is used to weight the group-to-point distance, which causes the shapes of the cluster hierarchy to be spherical or ellipsoidal \citep[see][Sect. 9.3.1]{Feigelson2012}. In this particular case, HC allows us to group stars that have 21 similar abundances (18 distinct elements, in which 3 are represented by both the neutral and ionised species), and thus 21 dimensions. This high-dimension clustering can only be achieved with techniques such as this one. HC also allows us to probe whether the concept  of CEF  \citep[][]{Boesso2018} is accurate for this particular problem. The results of the HC are displayed as a dendrogram, which can be seen in Fig. \ref{fig:dendrogram}.

It is worth mentioning that this technique is extremely useful to identify specific types of objects. For instance, in terms of chemical abundances, stars with defined chemical characteristics (e.g. solar, subsolar, and super-solar metallicities; metal-rich and  metal-poor; super-metal-rich and -poor) in a high-dimensional space. HC can also be used to constrain other stellar properties, such as orbital and/or atmospheric parameters. In the context of large sky surveys, the implementation of this simple yet powerful technique within various analyses pipelines can be advantageous in identifying objects with certain characteristics, and is not  limited to stars \citep[see e.g.][]{deSouzaCiardi2015, Sasdelli2016, deSouza2017}. In this paper we choose to cluster stars only based on their chemical abundances with the intent of probing their CEF and further analysing their orbital features.

For an initial analysis, the sample was divided into six main groups of stars using $d_W$ = 6. At this level we identified a group of 171 stars as the MMR group. With $d_W$=1.3, the MMR group can be further divided into five components, which we refer to as subgroups. We note that the $d_W$ numbers do not have a physical meaning, but are linked to the HC method. The $d_W$ values are arbitrary and were only chosen to optimise the visualisation of the groups and subgroups. The three different levels of groupings shown in Fig. \ref{fig:dendrogram} can be assessed in Table \ref{tab:hc_levels}. A summary of the MMR group and its five subgroups is given in Table \ref{tab:mmr_subgs}. It is important to note the small standard deviation in the distribution of the values of [Fe/H] and [Mg/Fe], which are of the order of 0.08 dex or less in each of the identified subgroups. This shows that we managed to divide the MMR group into subgroups that show a certain degree of chemical homogeneity, although outliers are still present.

Throughout this paper we concentrate on discussing the properties of the MMR group. For comparison, in some of the plots we also show the behaviour of a group with abundances closer to the solar values, hereafter called the `Solar group'.


\section{Results and  discussion} \label{sec:results}

\subsection{Chemical abundances}

\begin{figure*}
    \centering
    \includegraphics[width=\linewidth]{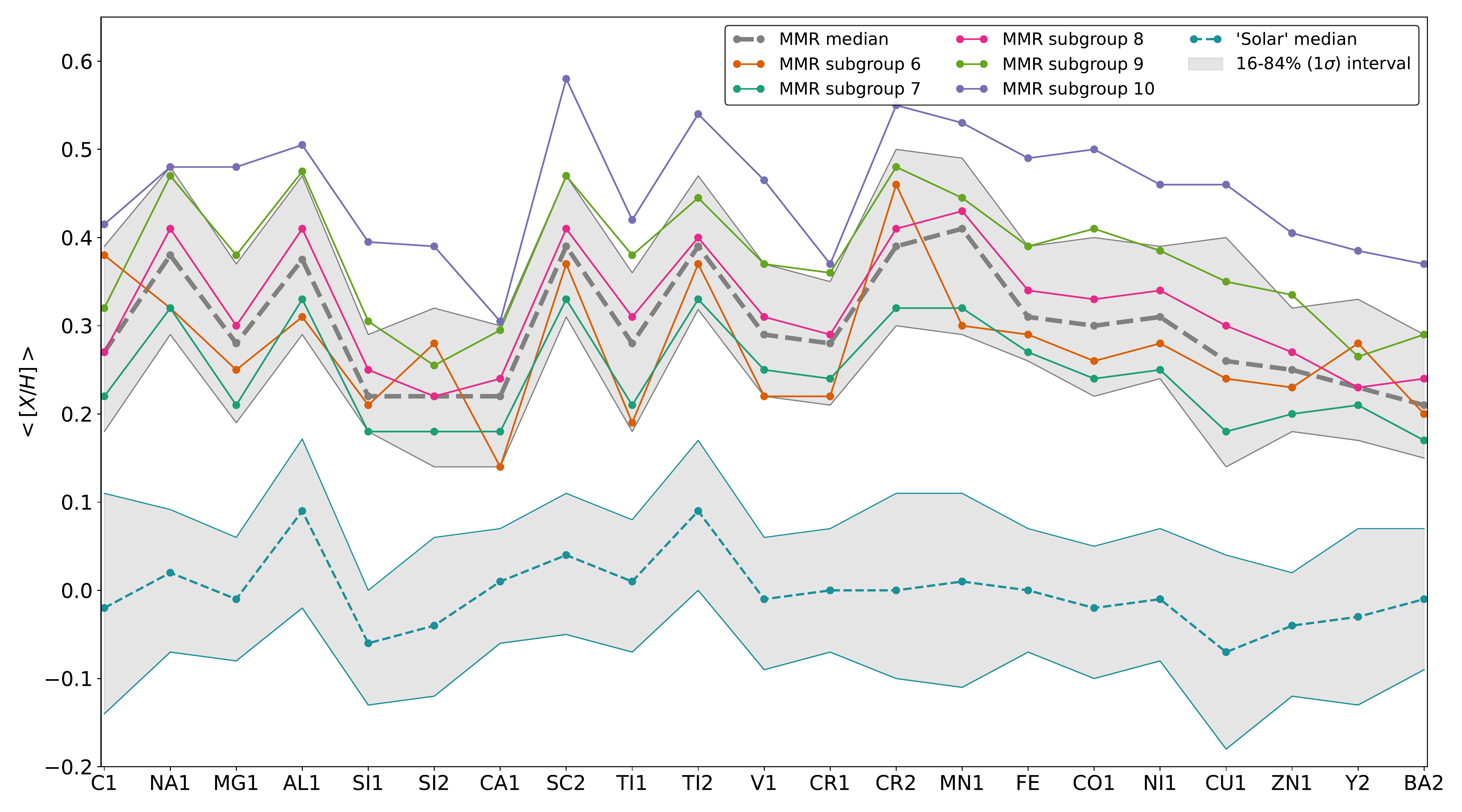}
    \includegraphics[width=\linewidth]{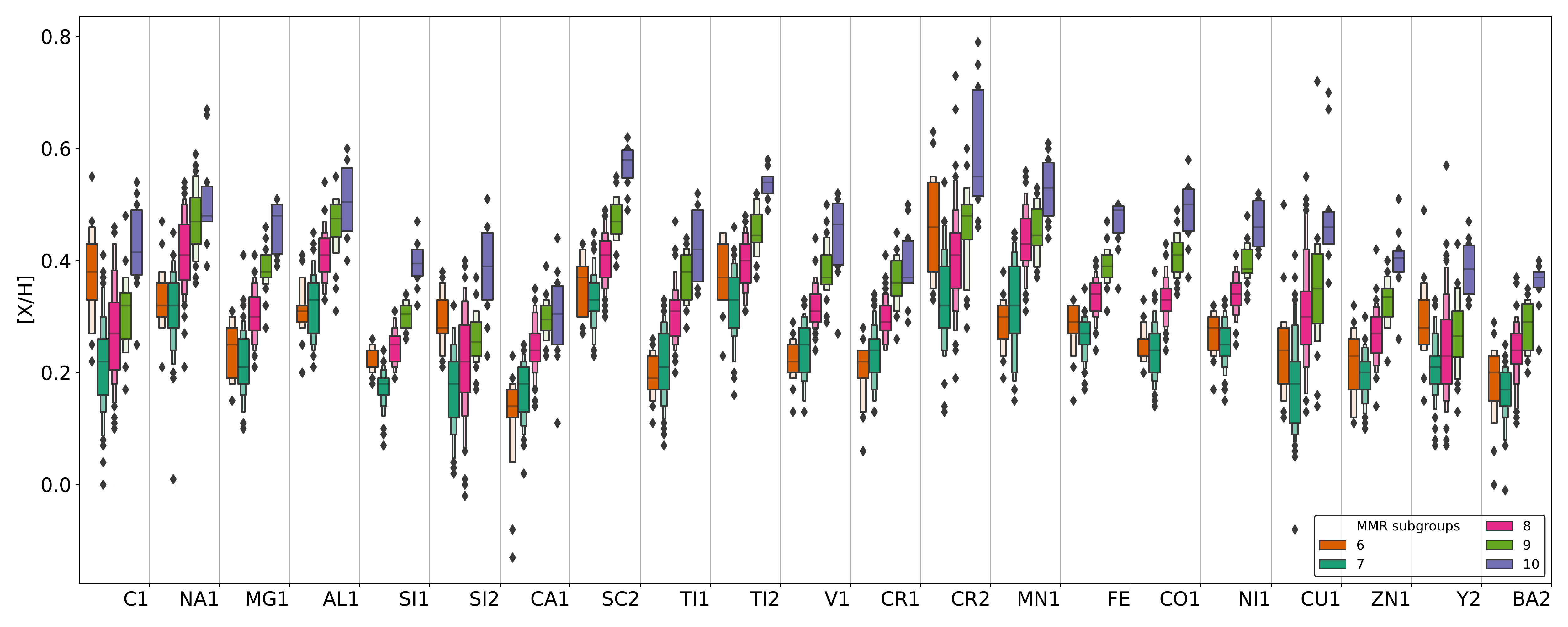}
    \caption{Distribution of the individual species used in the HC. \textit{Top panel:} Median abundances $\langle \rm{[X/H]} \rangle$ ($y$-axis) of each element X ($x$-axis) for the most metal-rich (MMR) group. The grey and cyan thick dashed lines are the median abundances of each element of the MMR and the Solar groups, respectively; the grey area represents the  1$\sigma$ (16--84\%) confidence interval for each one. The purple, light green, pink, dark green, and orange lines represent respectively the MMR subgroups 6 to 10 given by the hierarchical clustering in Fig. \ref{fig:dendrogram}. This image does not represent a regression; the lines connecting the abundances act as a visual guide to depict whether the abundance of each element increases or decreases in comparison with the previous one. \textit{Bottom panel:} Same as  top panel, but with the distributions of the same abundances ([X/H]) in the shape of letter-value plots \citep[a more general version of the boxplot,][]{Hofmann2017} with \texttt{k\_depth} set as \texttt{`trustworthy'}. The colours are the same as the top panel, but with the Solar group omitted. The vertical lines divide the distributions by element. The outliers are the diamond-shaped fliers. The bottom panel was made with \textsc{seaborn}, a \textsc{python} package for data visualisation \citep{Waskom2021}. In both panels the abundances are shown in increasing order of atomic number.}
    \label{fig:abund_lvplot}
\end{figure*}

In the top panel of Fig. \ref{fig:abund_lvplot} the median abundance pattern ($\langle \rm{[X/H]} \rangle$) of the MMR and  the Solar   groups are respectively depicted by the thick dashed grey and cyan lines; their 1$\sigma$ confidence interval is in grey and was estimated by re-sampling the data via bootstrap (1000 observations). The thin lines in orange, dark green, pink, light green, and purple respectively  represent the median abundances for the MMR subgroups 6 to 10.\footnote{The numbering was assigned in the joint analysis of the larger sample of 1460 stars. We keep it here for consistency.}  It is noticeable that the subgroups 7 to 10 (dark green, pink, light green, and purple), in this order, appear to be well stacked one upon the other, but group 6 (orange) does not. The abundance of each element seems to follow an increasing pattern, suggesting a well-ordered CEF, which we discuss further in Sect. \ref{subsec:ages}. However,  group 6 appears to differ from the others, and does not follow this pattern. It has higher median abundances for \ion{C}{i}, \ion{Si}{ii}, \ion{Sc}{ii}, \ion{Ti}{ii}, \ion{Cr}{ii}, and \ion{Y}{ii}, in comparison to the  group 7 (dark green), which  has a similar median [Fe/H]. 

Since these differences are mostly found in ionised species, we initially suspected problems in the determination of the atmospheric parameters of these stars, in particular an incorrect  value for the  surface gravity. However, we traced back the issues to incorrect abundances. Most of the spectral lines used to compute abundances for these species are weak and/or blended. Moreover, the majority of these lines are in the bluer part of the UVES spectrum (the arm covering from 480 to 580 nm), where the  \texttt{S/N} is actually below the limit of 40 that we used to select the best data. A few of these spectra are also affected by rotational broadening (with v$\sin{i}$ $\sim$ 8-9 km s$^{-1}$), although below our cutoff for this parameter. All these factors conspired together to produce unreliable abundances. Therefore, we conclude that subgroup 6 does not possess any real chemical peculiarity, but joins a series of stars affected by analysis problems that influence the abundances of these few species.  

We also decided to analyse individually a few of the upper outliers in the abundances of \ion{Cu}{i} with the goal of confirming whether these systems were bona fide Cu-rich stars. We also found that there were problems in the estimations of this abundance for these outliers (the three points with high [Cu/H] in the bottom panel of Fig.~\ref{fig:abund_lvplot}). We recommend that outliers such as these be treated cautiously as they may lead to unwarranted conclusions.

\begin{figure*}
    \centering
    \includegraphics[width=\linewidth]{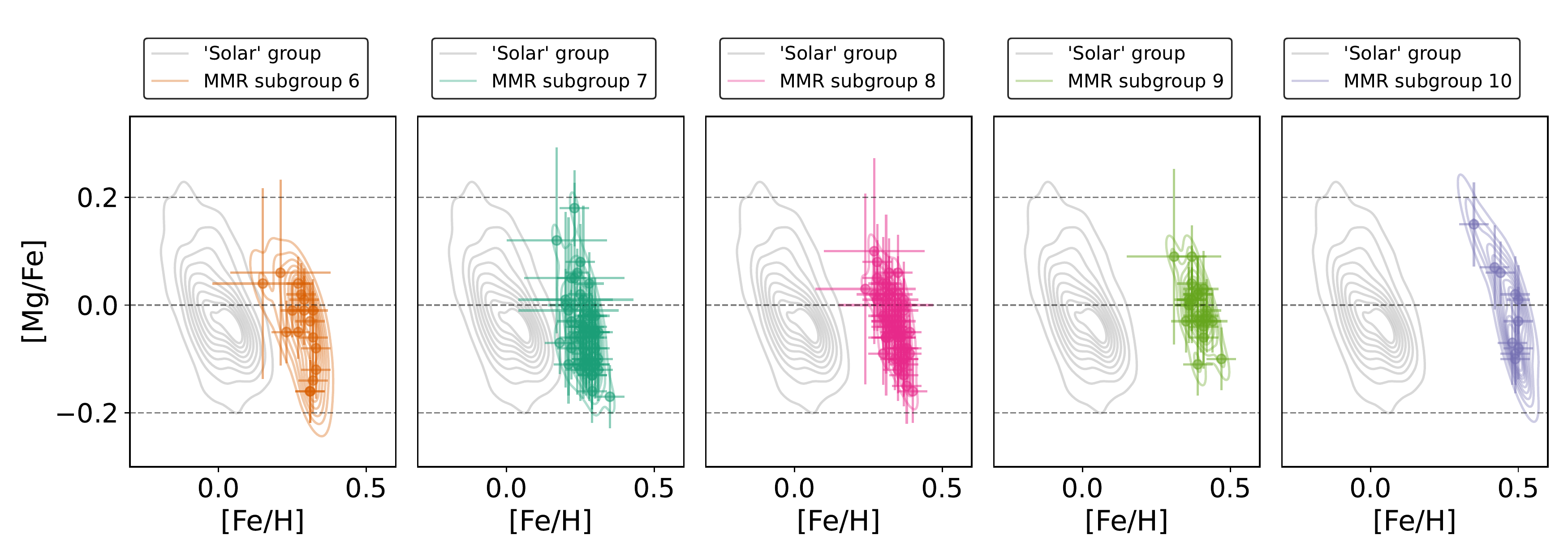}
    \caption{[Mg/Fe] {vs} [Fe/H] for all stars in each MMR. The Gaussian kernel densities are also displayed along with those from the Solar group.}
    \label{fig:feh_mgfe}
\end{figure*}

All the metal-rich groups have low ratios of $\alpha$-elements to iron ([$\alpha$/Fe]), as displayed in Fig.~\ref{fig:feh_mgfe}. This is an indication of a significant contribution from supernovae (SN) type Ia to the material from which the stars formed. Moreover, the abundances of carbon relative to iron, [C/Fe], with a median within $-$0.1 and 0.0 dex, suggest a contribution from asymptotic giant branch stars, as otherwise lower carbon would be expected \citep{Kobayashi2020}. Overall, we conclude that the abundances indicate that these stars originated from a region of complex chemical enrichment. In particular, we note here that old stars from the bulge display high ratios of [$\alpha$/Fe] that eventually decrease with increasing [Fe/H], reaching low [$\alpha$/Fe] values at metallicities similar to those of our stars (see \citealt[][their Fig. 7,]{Hill2011} and \citealt[][their Fig. 20)]{Bensby2017}. The values of [Mg/Fe] for the entire sample of super-metal-rich stars roughly vary between 0.2 and $-$0.2, which is in agreement with what is seen in the APOGEE sample for metal-rich stars with $Z_{\rm{max}}$ similar to ours \citep[between $\sim$0.5--1.5 kpc; see][]{Hayden2015}.

All the MMR subgroups have a distribution of [Mg/Fe] that is not too different from that of the Solar group (also shown in Fig. \ref{fig:feh_mgfe}, in grey). Low [$\alpha$/Fe] is a characteristic associated with thin disc membership, as opposed to thick disc stars which display high [$\alpha$/Fe] \citep[e.g.][and references therein]{RB2014}. There is some discussion in the literature on whether the thick disc extends to high values of metallicity \citep{Hayden2017, Lagarde2021}. At high metallicities, a possible thin versus thick disc separation in terms [$\alpha$/Fe] is never as clear as it is at low metallicities \citep[as seen in e.g.][]{Fuhrmann2021}. In this work, we preferred   not to attempt to assign the MMR stars to the thin or thick disc. We just note that, based on a chemical criterion, the general low values of [Mg/Fe] would classify most of these stars as belonging to the thin disc, even in works where a metal-rich $\alpha$-enhanced population is defined \citep{Hayden2017, Lagarde2021}. 

\subsection{Ages} \label{subsec:ages}

\begin{figure*}
    \centering
    \includegraphics[width=\linewidth, trim={0 0.5cm 0 0}, clip]{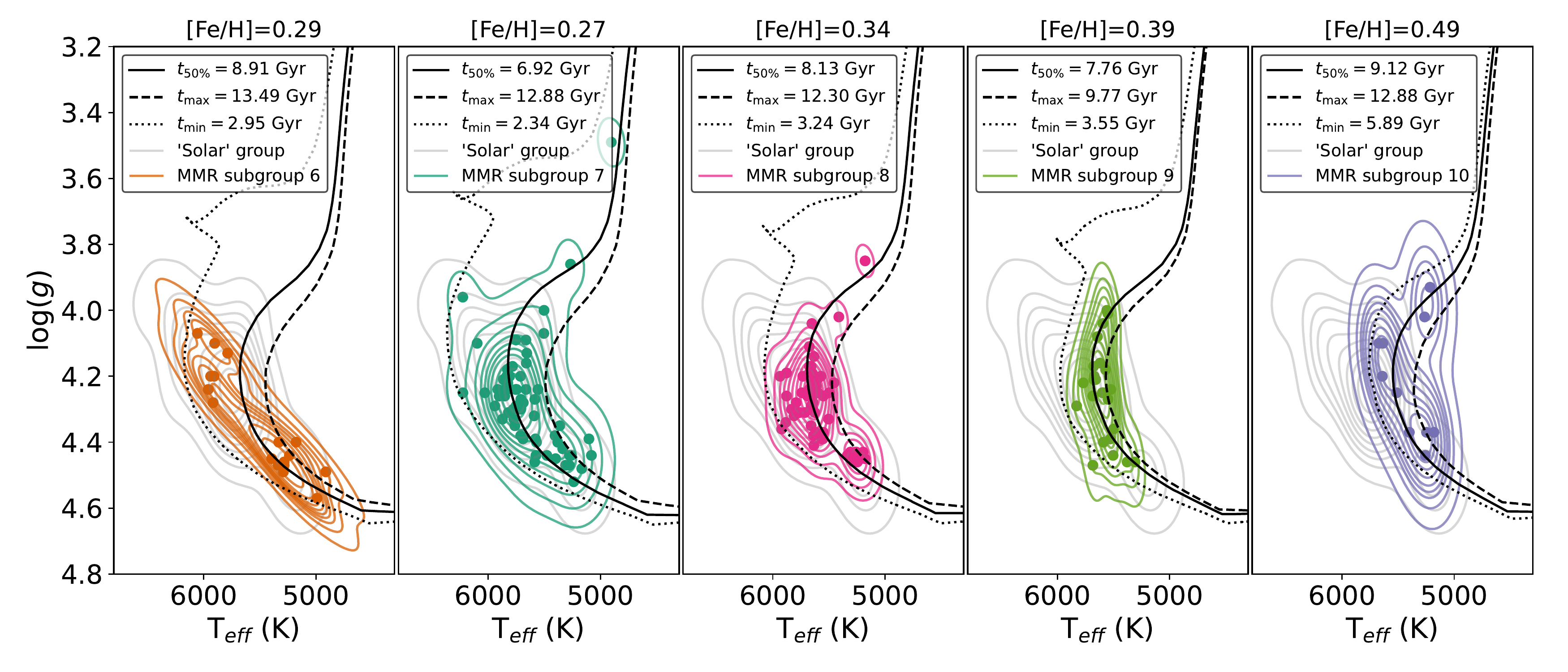}
    \includegraphics[width=\linewidth, trim={0 0 0 1cm}, clip]{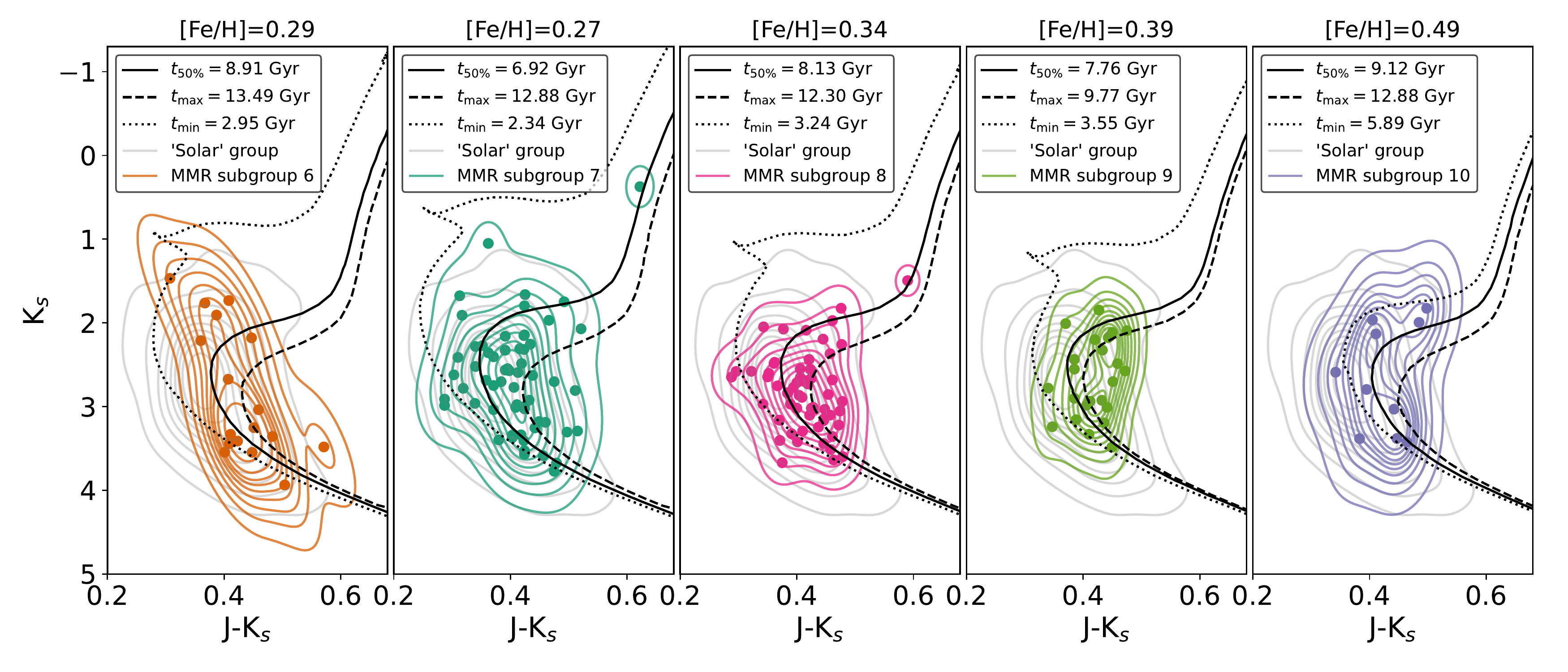}
    \caption{Kiel and colour-magnitude diagrams for each of the MMR subgroups. \emph{Top panel:} Surface gravity \emph{\emph{vs}} effective temperature [$\log(g)$ \emph{\emph{vs}} $T_{\rm{eff}}$] for the five MMR subgroups, in the same colours as Fig. \ref{fig:abund_lvplot}, in the shape of 2D kernel density plots with their respective scatter markers. In grey we display the Solar group parameters for comparison. Isochrones for the maximum, median, and minimum ages (according to Table \ref{tab:mmr_subgs_ages}) are also displayed as dashed, straight, and dotted lines, respectively. All isochrones shown are in terms of the median values of [Fe/H] shown in Table \ref{tab:mmr_subgs}. The stars  shown  have a distribution of metallicities that are not shown in the isochrones. \emph{Bottom panel}: Equivalent CMD of the top panel. The CMD uses \emph{JHK$_s$} absolute magnitudes from 2MASS, which were used by \textsc{unidam} to estimate the ages.}
    \label{fig:logg_teff_kde}
\end{figure*}

In Fig. \ref{fig:logg_teff_kde}, we show diagrams of effective temperature {versus} surface gravity ($T_{\rm{eff}}$ \emph{\emph{vs}} $\log(g)$) in the upper row, and colour-magnitude diagrams (CMDs, with 2MASS magnitudes), in the lower row. Both types of diagrams in Fig. \ref{fig:logg_teff_kde} are displayed along with the isochrones generated with the parameters listed in Table \ref{tab:mmr_subgs}. The distribution of ages in all MMR subgroups can also be assessed in Fig. \ref{fig:appendix_agedistribution}. All the MMR subgroups have median ages $\overline{t}$ between $\sim$7--9 Gyr, but with a certain variety of median metallicities ($\langle \rm{[Fe/H]} \rangle$ ranging between +0.27 and +0.49 dex). The PARSEC isochrones \citep{Bressan2012} shown  were retrieved from CMD 3.6 web interface.\footnote{\url{http://stev.oapd.inaf.it/cgi-bin/cmd}} The bottom row in Fig. \ref{fig:logg_teff_kde} seems to depict better isochrone fits, as \textsc{unidam} uses such magnitudes to fit the PARSEC isochrones. Five of the stars in the MMR group (two for subgroup 6 and three for subgroup 7) have very low age estimations ($\sim 10$ Myr) in contrast to their median ages of $\sim$ 7--9 Gyr. Since these stars have the same characteristics overall of the whole group, except by being at the lower limits of the  isochrone turnoff (consequently having a lower age estimation), we speculate whether they are blue stragglers. 

Blue stragglers are objects that can originate from mass transfer in binary  (or even trinary) systems or from stellar mergers, and appear to be younger than they really are \citep[e.g.][and references therein]{Perets2009, Davies2015}. This type of star seems to be more common than previously thought and, in fact, they populate all the regions of the MW \citep{Santucci2015}. Unlike \citet{Chen2019}, who identify two groups of stars, one with  stars  mostly older than 4 Gyr and another with stars younger than  1 Gyr, in our sample we do not identify two groups of different ages. This is the case, since the young stars of \citet{Chen2019} have $T_{\rm{eff}}$ $>$ 7000 K, a regime that is not included in our sample. In our case, most stars are estimated to be old (with a peak at $\sim$ 7.76 Gyr) with five   much younger outliers ($\sim$ 10 Myr). 

At the solar neighbourhood,  ages like those observed here are usually seen in thick disc (or even halo) stars. Thick disc stars are generally older than $\sim$8 Gyr, while thin disc stars are for the most part younger than that \citep{Haywood2013}. The ages alone would then imply that most of our sample is made of thick disc stars, which seems incompatible with their chemical abundances (i.e. high metallicities and low [Mg/Fe] ratios). However, we note that the typical uncertainty in age can be around 1--2 Gyr (see the  discussion on the precision of the ages derived with \textsc{unidam} in \citealt{Mints2018}), which would push $\overline{t}_{50\%}$ (as shown in Table \ref{tab:mmr_subgs_ages}) to 4.72--7.72 Gyr at the lower end and to  8.92--11.01 at the upper limit. The age range still agrees better with that of thick disc stars, but the presence of a certain fraction of thin disc stars cannot be excluded. 

It is worth mentioning that the isochrones shown in Fig. \ref{fig:logg_teff_kde}, for visualisation purposes, take into account only the median values of [Fe/H] and three ages (maximum, minimum, and median), excluding the outliers of younger age, which we speculate might be blue stragglers.  Each MMR subgroup still displays a certain range in [Fe/H], even after being clustered via HC. The real values of [Fe/H] for these stars are based on a distribution described in Table \ref{tab:mmr_subgs}.

The distribution of chemical abundances shown in Fig. \ref{fig:abund_lvplot} combined with the age estimations shown in Fig. \ref{fig:logg_teff_kde} does not seem to support the idea that the subgroups are all part of a single CEF. For the CEF to be true, in the most straightforward case the sequence of increasingly metal-rich subgroups should also be a sequence of decreasing median age (a more metal-rich subgroup being formed out of material enriched by the previous  older  subgroup). On the contrary, we find that the most metal-rich subgroup (number 10) is also the  one that contains the oldest stars. The next subgroup with the oldest median age (number 6) is less metal-rich than subgroup 10. This in itself suggests that, despite their similar ages, the places where these subgroups of stars were formed must be distinct. Therefore, our analysis points out that not only these subgroups of stars originated from the inner regions of the MW, but also in different radii.

\subsection{Chemo-dynamic features}

To probe in more detail the potential origins of the stars in the MMR subgroups, we now include in the discussion the parameters generated by \textsc{galpy} (see Sect. \ref{subsec:orbits_age}) in combination with the abundances and metallicity, and analyse their properties. 

\subsubsection{Orbital features and signatures of migration}

\begin{figure*}
    \centering
    \includegraphics[width=\linewidth]{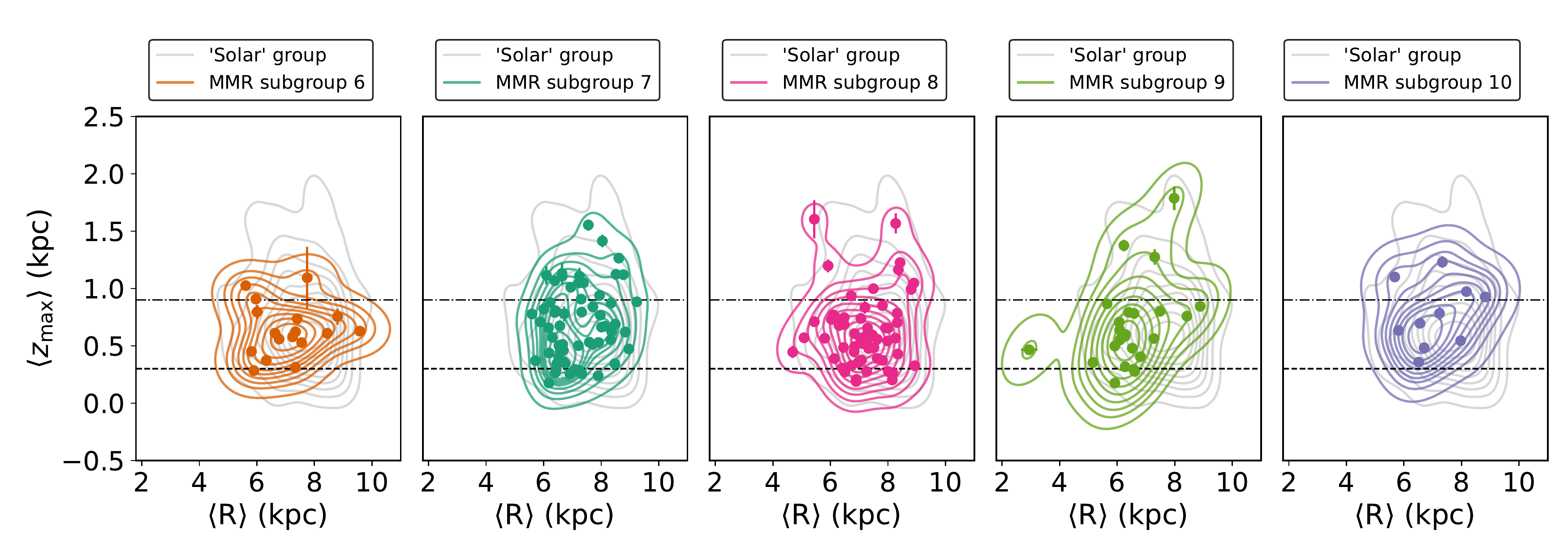}
    \caption{Median highest position in the plane of the MW \emph{\emph{vs}} median guiding radius, both in kpc, ($\langle Z_{\rm{max}} \rangle$ \emph{\emph{vs}} $\langle \rm{R} \rangle$) for the five MMR subgroups (in the same colours as Fig. \ref{fig:abund_lvplot}) in the shape of 2D kernel density plots with their respective scatter markers. The Solar group parameters are shown in grey for comparison. No star has negative $\langle Z_{\rm{max}} \rangle$, but  the images have been enlarged up to $\langle Z_{\rm{max}} \rangle=-0.5$ for better visualisation of the density curves. As a visual guide, the thin and thick discs thresholds are shown as black dashed and dot-dashed lines, respectively.}
    \label{fig:zmax_rguiding_kde}
\end{figure*}

\begin{figure*}
    \centering
    \includegraphics[width=\linewidth]{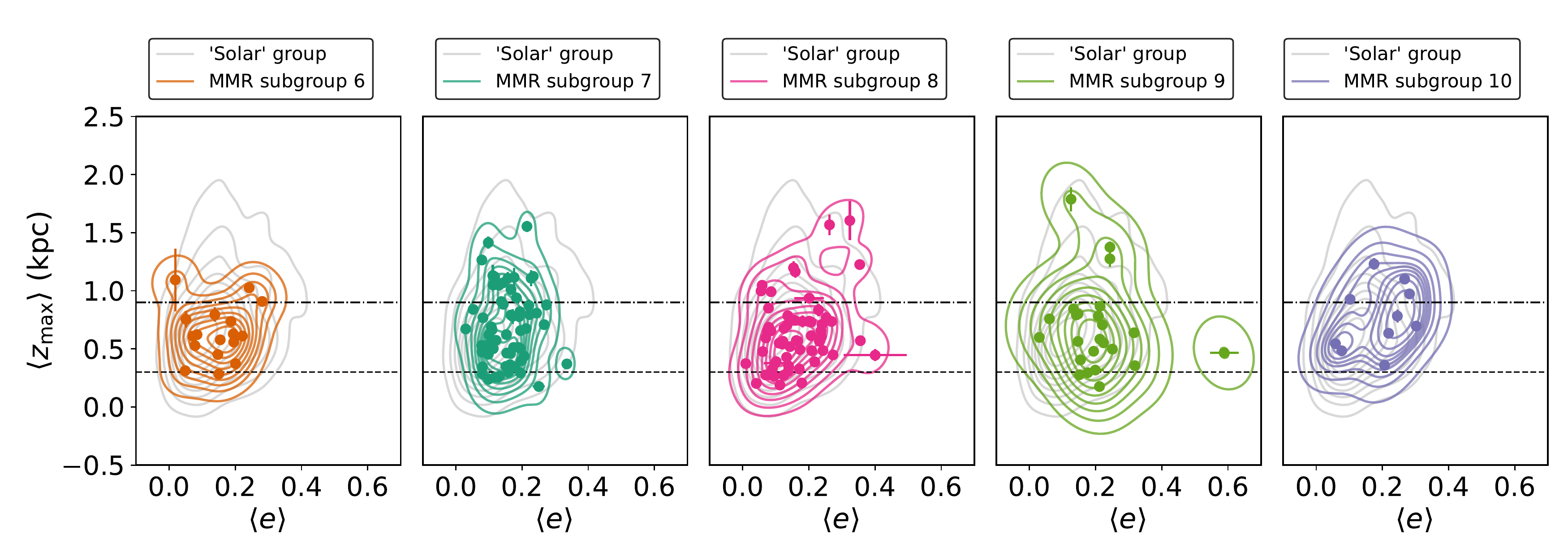}
    \caption{Median highest position in the $z$-axis of the MW \emph{\emph{vs}} median eccentricity, both in kpc ($\langle Z_{\rm{max}} \rangle$ \emph{\emph{vs}} $\langle e \rangle$) for the five MMR subgroups (in the same colours as in Fig. \ref{fig:abund_lvplot}) in the shape of 2D kernel density plots with their respective scatter markers. The Solar group parameters are shown in grey for comparison. No star has $\langle Z_{\rm{max}} \rangle$ or $\langle e \rangle$<$0$, but the images have been enlarged for better visualisation of the density curves. As a visual guide, the thin and thick disc thresholds are shown as  black dashed and dot-dashed lines respectively.}
    \label{fig:zmax_ecc}
\end{figure*}

Figures \ref{fig:zmax_rguiding_kde} and \ref{fig:zmax_ecc} display, respectively, the median highest distance from the MW plane ($\langle Z_{\rm{max}} \rangle$) against the median guiding radius ($\langle R \rangle$), and $\langle Z_{\rm{max}} \rangle$ against the median eccentricity ($\langle e \rangle$) for each star in each MMR subgroup. In both figures, the scale heights of the thin and thick discs are displayed by the dashed and dot-dashed lines, respectively \citep[e.g.][]{McMillan2017} (see Table \ref{tab:mw_properties} in the Appendix). It is possible to see that most stars in all the MMR subgroups reach values of $Z_{\rm{max}}$ where a significant population of thick disc stars is expected ($0.3 \leq \langle Z_{\rm{max}} \rangle\,\mathrm{(kpc)} \leq 0.9$). Very few stars are found with $\langle Z_{\rm{max}} \rangle \leq 0.3$ kpc, where the thin disc would be clearly dominant. Overall, at first glance, the values of $Z_{\rm{max}}$ would suggest that the MMR stars could be a mix of the different components of the MW disc. 

\citet{Roskar2013} show via numerical simulations that a population of stars that migrates towards the outskirts of the Galaxy, because of perturbations induced by the spiral structure, causes the thickening of the Galactic disc. For instance, they show that stars of $\sim$8--9 Gyr, formed in a radius within 2$<$R (kpc)$<$4, can have an increase in $\langle Z_{\rm{max}} \rangle$ of 0.9 kpc (see their Fig. 2). The authors also argue that the simulations are made in a setting without cosmological perturbations and add that the thickening of the disc should be increased in a more realistic scenario, which is the case for the MW. Thus, qualitatively at least, the $Z_{\rm{max}}$ distribution of our stars seems to agree with the idea that they were formed in the inner disc and with smaller $Z_{\rm{max}}$ values, reaching their current properties due to migration. Nevertheless, there are other works that argue the exact opposite, that radial migration has a little impact on disc thickening \citep[e.g.][]{Minchev2012b, Halle2015}.

The vertical velocity dispersion ($\sigma_{v_z}$) is another parameter that might help to reveal whether stars have migrated from the inner Galaxy \citep{Halle2015}. In Table \ref{tab:vel_disp} we display the results for the velocity dispersion in $z$, $r$, and $\phi$ for all MMR subgroups. According to \citet{Halle2015}, it is possible to identify stars that originate in the inner disc and have been subject to radial migration, as they are those with the highest values of $\sigma_{v_z}$ between stars of similar age and radii (by $\sim$ 20\% in general and up to 50\% higher for the most extreme cases). For comparison, typical values for $\sigma_{v_z}$ can be estimated via the relation given by \citet{Sharma2021}. In our sample, subgroup 10 (purple) has the highest value of $\sigma_{v_z}$ (23.99 km s$^{-1}$). For the metallicity of this subgroup ([Fe/H] $\sim$ +0.49), a typical $z$ height of about 0.5 kpc, median age of $\sim$ 9 Gyr, and typical values of angular momentum seen in our sample, the \citet{Sharma2021} relation returns values of $\sigma_{v_z}$ between 16 and 20 km s$^{-1}$. Therefore, the $\sigma_{v_z}$ of subgroup 10 is  higher by a factor between 20--50\%, indicating that these stars are likely extreme migrators. For a lower metallicity ([Fe/H] = +0.3) and the same values for the other quantities as above, the \citet{Sharma2021} relation returns values of $\sigma_{v_z}$ between 18.5 and 22.6 km s$^{-1}$ (and in the range 16.8--20.5 km s$^{-1}$ at the Galactic plane). The values of $\sigma_{v_z}$ for the other subgroups are within the expected ranges. The evidence for migration from the vertical velocity dispersion of these subgroups is thus not clear, but the possibility cannot be fully excluded. The typical ranges quoted above are wide enough to hide a variation of $\sim$ 20\% expected from the work of \citet{Halle2015}.

Additionally, when assessing the velocity dispersion of the stars in the solar vicinity, many factors (such as statistics) can influence the final results, including the volume occupied by these stars (i.e. the larger the radii, the more varied the stars) and the number of stars considered. This is probably why subgroup 10 (purple) has such a large velocity dispersion in the $z$ direction: this group is composed of only ten objects (see Table \ref{tab:mmr_subgs}). The orange subgroup (number 6) is the second smallest subgroup and also the second in the order of largest $\sigma_{v_z}$, while subgroups 8 and 9 (pink and light green, respectively) are the most numerous and also possess the lowest values of $\sigma_{v_z}$. All in all, if we consider these biases, it is very hard to make any strong assumptions in terms of migration for our sample only looking at $\sigma_{v_z}$. In terms of the volume they occupy, as previously mentioned, they are all within $\sim 2$ kpc of the Sun, which can be seen in Fig. \ref{fig:heliocentric_distances}, where  the distances are projected in the $xy$ and $xz$ Galactic heliocentric coordinates. 

Furthermore, the distribution of eccentricities in our sample is in agreement with the predictions of radial migration induced by the bar as modelled by \citet{Khoperskov2020}. According to their \emph{N}-body simulations of the MW, one would expect migration to circularise (i.e. decrease in eccentricity) the orbit of metal-rich stars from the inner disc that reach the solar neighbourhood. In addition, according to \cite{Roskar2008, Roskar2012}, who studied the effects of the spiral structure on radial migration, significant changes in the orbital eccentricities are not expected. These results agree with what we observe in our study: super-metal-rich stars with low eccentricity values, $e$ $\lesssim$ 0.2 \citep[as also found by][]{Kordopatis2015, Hayden2015}. Moreover, the low eccentricities suggest that churning, and not blurring, might be the dominant effect behind the motion of the stars in our sample, although we   note that the two effects are probably acting together.

Altogether, considering the several parameters  presented here (e.g. ages, metallicities, eccentricities, maximum vertical scaleheights), it is likely that our stars were drawn from the inner Galaxy by a combination of churning and blurring, and were not a part of a peculiar thick disc-like stellar population with thin disc-like metallicities.

\subsubsection{Radial metallicity profile}

\begin{figure*}
    \centering
    \includegraphics[width=\linewidth, trim={0 0.6cm 0 0}, clip]{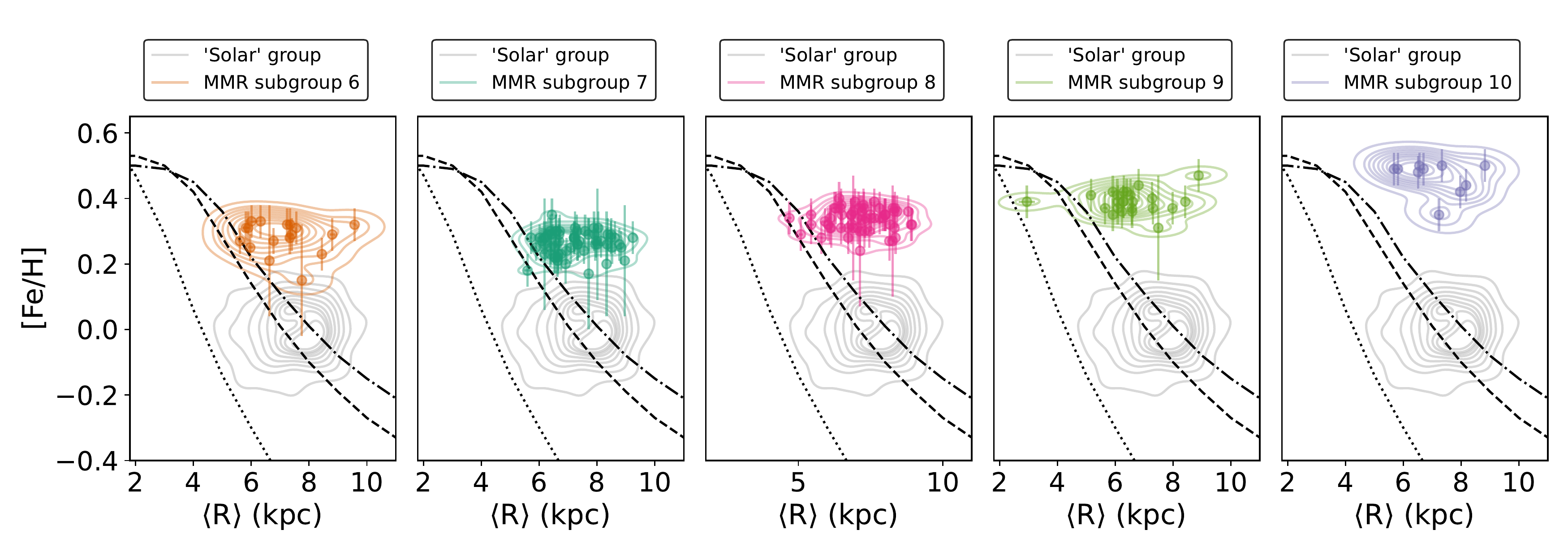}
    \includegraphics[width=\linewidth, trim={0 0.6cm 0 2.4cm}, clip]{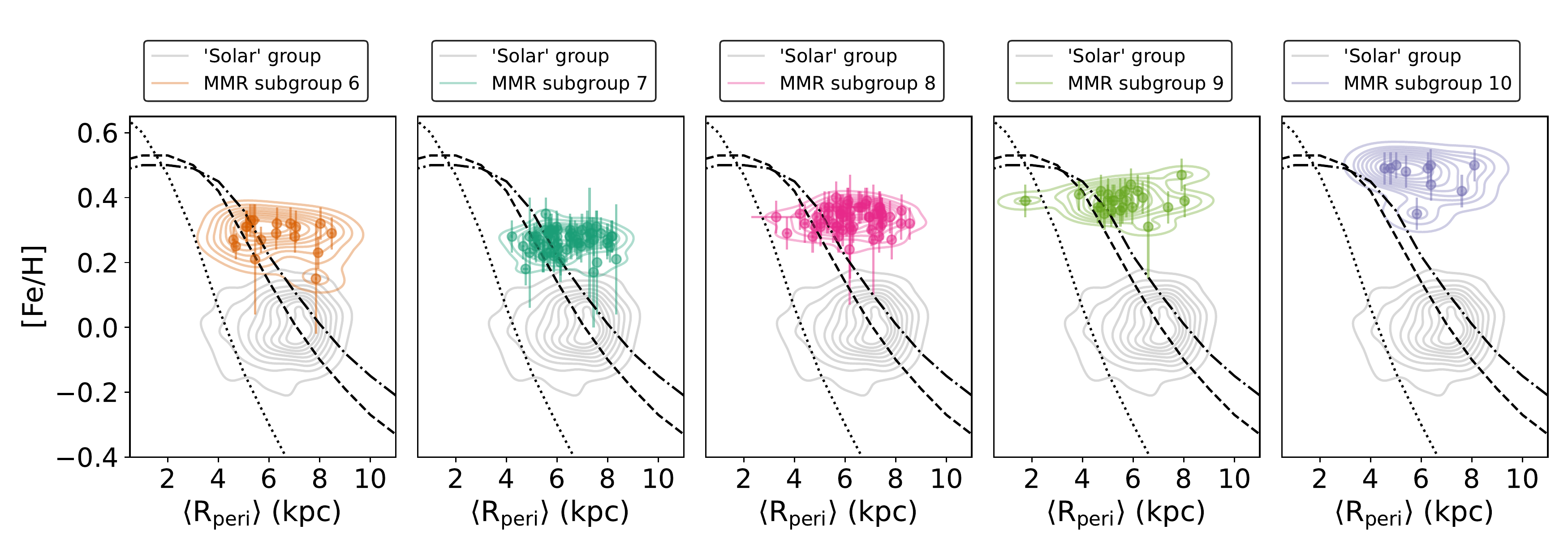}
    \includegraphics[width=\linewidth, trim={0 0.6cm 0 2.4cm}, clip]{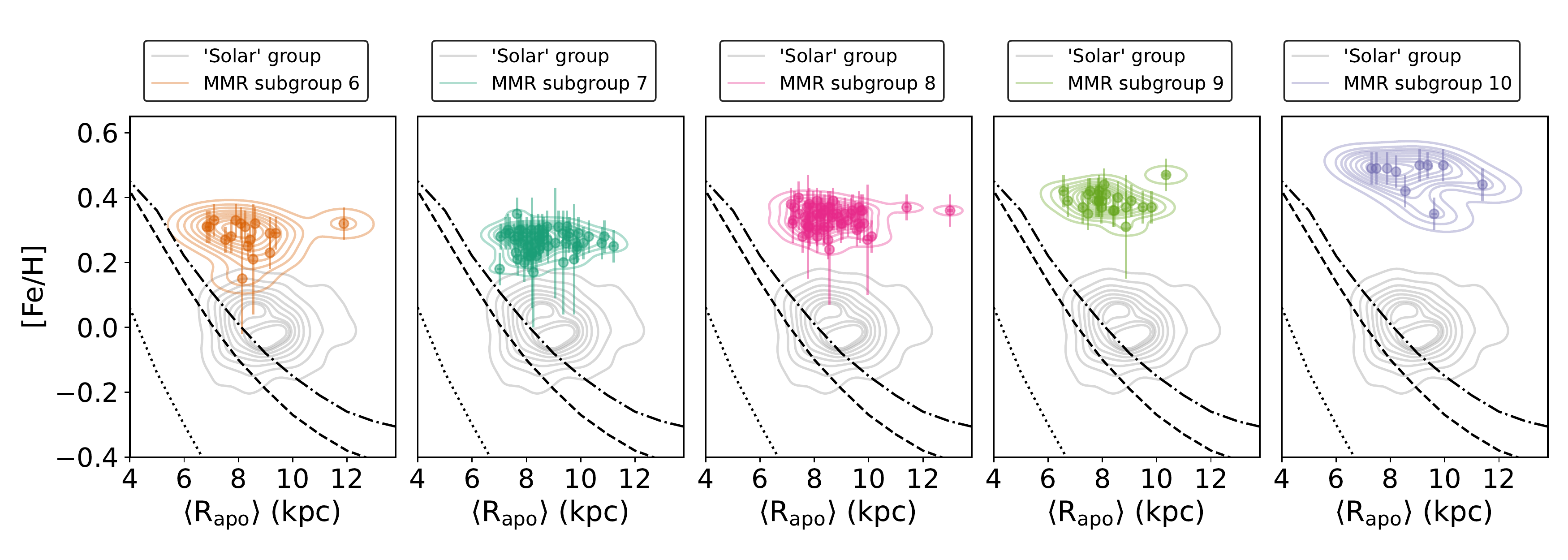}
    \caption{[Fe/H] \emph{\emph{vs}} $\langle \rm{R} \rangle$ (top), $\langle \rm{R_\textrm{peri}} \rangle$ (middle), and $\langle \rm{R_\textrm{apo}} \rangle$ (bottom) for all stars in each MMR subgroup; their 2D-Gaussian kernel densities are also displayed along with those from the Solar group. The dotted, dashed, and dot-dashed black curves respectively depict the 3.3, 8, and 11 Gyr models described in \citet{Magrini2009}. In all MMR subgroups, the values of [Fe/H] are very high even at large radii. Note that for each case ($\langle \rm{R} \rangle$, $\langle \rm{R_\textrm{peri}} \rangle$, and $\langle \rm{R_\textrm{apo}} \rangle$) the $x$-axis range is not the same to enable a better visualisation of the groups.}
    \label{fig:feh_rguiding}
\end{figure*}

Figure \ref{fig:feh_rguiding} displays the values of [Fe/H] in terms of the orbital guiding radius $\langle \rm{R} \rangle$, the pericentric distance $\langle \rm{R_\textrm{peri}} \rangle$, and the apocentric distance $\langle \rm{R_\textrm{apo}} \rangle$, for all MMR subgroups and for the Solar group. These values are compared to the Galactic chemical evolution models at ages of 3.3, 8, and 11 Gyr from \citet{Magrini2009}, which are depicted by the dotted, dashed, and dot-dashed black lines in all the subplots, respectively.  The ages of the models show how the radial metallicity distribution of the MW should be at different ages of the Universe (i.e. when it was 3.3, 8, and 11 Gyr). The models at 8 and 11 Gyr are displayed as a guide to the state of the metallicity gradient around the time when the Sun was formed and thus as a good comparison to the Solar group. Since the Sun is about 4.7 Gyr old, it should be compared to a model curve at 9 Gyr, which is not available. The model at 3.3 Gyr (adequate for stars with an age of about 10.5 Gyr) is much steeper and is the one closer to the median ages of the MMR stars. The models display a known phenomenon, which is that the inner regions of the MW are richer in metals when compared to the outskirts of the Galaxy \citep[e.g.][]{FrancoisMatteucci1993, Grenon1987, Wilson1994, Grenon1999, Andrews2017}. This phenomenon does not appear to be unique to the MW; it happens in other systems as well \citep[e.g.][]{Henry1999}. In other words, this metallicity gradient phenomenon supports the inside-out galaxy formation scenario \citep[e.g.][and references therein]{Magrini2009, Bergemann2014, Andrews2017}. In all the MMR subgroups the stars are very different from the expectations of the 3.3 Gyr model. They are richer in metals for the estimated $\langle \rm{R} \rangle$ when compared to what the models predict. This result seems to support the idea,  at least in part, that these stars may have been formed in inner regions of the MW (such as the bulge or the inner disc) and then gradually moved to its outskirts due to churning and blurring. The fact that these metal-rich stars migrated to their current position seems to be in line with the suggestion of other works showing that a flattening of the radial metallicity distribution for old stars can be caused by radial migration (e.g. \citealt{Schonrich2009b, Minchev2013, Minchev2014, Grand2015, Kubryk2015_01, Kubryk2015_02}).

As seen in Fig. \ref{fig:feh_rguiding}, all the MMR subgroups have an extended radial distribution. Regardless of the median [Fe/H], the stars in each MMR span a range of at least 4 kpc, mostly between 6 and 10 kpc in guiding radius. If these stars were born with properties following an initial metallicity gradient, that gradient was not kept intact in their movement outwards from the inner disc. The distribution of these stars can be contrasted with that of the Solar group (which has a better alignment between the centre of its kernels) and the models from \citet{Magrini2009}; the Solar group stars seems more consistent with the metallicity gradient. The different median [Fe/H] of each MMR subgroup seems to imply that each group may have been formed in different radii from the Galactic centre. For instance, by comparing the distributions of [Fe/H] and the models, the stars in subgroup 7 seem to have been formed at R$\sim$4, whereas some of those in subgroup 10 appear to have been formed at R$\sim$2. This confirms what was discussed above, that subgroup 10 likely contains the stars that migrated the most through the disc, and for them churning seems to be the dominant process leading to migration. When their distributions of age and metallicity are considered, the other subgroups can be explained by a combination of blurring and churning. The comparison of the models with $\langle \rm{R_\textrm{peri}} \rangle$ indeed suggests that a fraction of the stars could have been formed in inner regions that agree with their metallicities. However, these stars still need churning to explain their current orbits. All in all, not only do our super-metal-rich stars seem to have been formed in the inner Galaxy, but also at different radii.  This hypothesis is also supported by the apparent lack of a well-defined CEF in our analysis, as discussed in Sect. \ref{subsec:ages}.


\subsubsection{Additional discussion}

Since these super-metal-rich stars are quite old ($\sim$ 7.76 Gyr;  see Table \ref{tab:mmr_subgs_ages}) they were susceptible to all the interactions the Galaxy suffered throughout their lifetime. This would also explain the disturbance in their orbits, making them reach larger heights in the plane of the MW, as seen in Figs. \ref{fig:zmax_rguiding_kde} and \ref{fig:zmax_ecc}. These results seem to be even more important for subgroup 10 (in purple), which is the most rich in metals of all the MMR subgroups (which can also be seen in Fig. \ref{fig:feh_mgfe}); it is possible that such stars could have come from even more internal regions of the MW. Nevertheless, the estimated eccentricities fall into the low regime; in other words, nearly all stars in all MMR subgroups have $\langle e \rangle < 0.4$, being more concentrated up to $\langle e \rangle \sim 0.2$ (see Fig. \ref{fig:appendix_eccentricities}). Only one star in subgroup 9 appears to have a high  eccentricity ($\langle e \rangle \sim 0.6$), and the median $\langle e \rangle$ for this subgroup seems to be just a little above 0.2. These results are consistent with the findings of \citet[][see their Fig. 7]{Hayden2018}. The idea of an inner disc origin for such metal-rich stars has been  discussed at length in the works of M. Grenon and collaborators \citep[e.g.][]{Grenon1987, Grenon1999, Pompeia2003, Trevisan2011}, and our results in fact match this hypothesis.

These findings are in agreement with the idea that, in fact, various dynamical processes (such as blurring and churning) can cause the mixing of the stars. Our MMR sample seems to have been mostly influenced by churning, which is the actual process most works equate to radial migration \citep{SellwoodBinney2002}, although we cannot exclude that for a few stars in our sample blurring has also been important as a mechanism to give the stars their current position. Our results are in agreement with several other works in the literature that made use of data from large spectroscopic surveys, such as \citeauthor{Hayden2015} (\citeyear{Hayden2015, Hayden2018}), \citet{Kordopatis2015}, \citet{Chen2019}, and  \citet{Khoperskov2020}, to mention a few. \citet{Hayden2018} also made use of the \textit{Gaia}-ESO survey to investigate radial migration in metal-rich stars, but using a different sample observed with the GIRAFFE spectrograph at lower resolution, while \citet{Kordopatis2015} used data from the Data Release 4 of the RAdial Velocity Experiment \citep[RAVE;][]{Steinmetz2006, Kordopatis2013}. It is also worth mentioning that \citet{Adibekyan2011} found two populations of metal-rich and metal-poor $\alpha$-enhanced stars from both the thin and thick discs, and suggest that they might have   originated in the inner Galactic disc. Our results differ from theirs as we found a population of super-metal-rich stars that outwardly seemed to share characteristics of both the thin and thick discs, but  upon further inspection,  by combining all their chemo-dynamic characteristics, they were found to most likely originate in the inner Galaxy.

Additional supporting plots and tables are found in the Appendix. As previously mentioned, Fig. \ref{fig:heliocentric_distances} shows the spatial location of the stars in our sample with their heliocentric positions. Figure \ref{fig:appendix_agedistribution} depicts the age distributions for the stars in each MMR group; the left panel shows the entire sample distribution with histograms and 1D Gaussian kernel densities and the right panel zooms into $8.5 \leq \log(t) \leq 11.5$ displaying the age medians. In Fig. \ref{fig:appendix_eccentricities} it is possible to see the distributions of $\langle e \rangle$ for all the MMR subgroups, with all the groups with medians near to or lower than 0.2.  Figure \ref{fig:appendix_toomre}  displays the Toomre diagram for all the stars in the MMR group of stars corrected for the Sun's velocity. It is possible to see that all stars are moving at different velocities, but all with prograde rotation with the disc. Figure \ref{fig:appendix_energy_lz} depicts the Lindblad diagram for all the subgroups, showing that all the stars posses a prograde orbit, with different levels of binding energy. Table \ref{tab:vel_disp} displays the velocity dispersion for the stars in all MMR subgroups in cylindrical coordinates.

\section{Summary, conclusions, and final remarks} \label{sec:conclusions}

We selected a group of 171 of the most (super-)metal-rich stars in the \textit{Gaia}-ESO iDR6 survey by grouping the stars by similarity of chemical abundances via hierarchical clustering. With this sample, we analysed its chemo-dynamic properties. Our findings, conclusions, and final remarks are as follows:

\begin{enumerate}
    \item Outliers in the distribution of abundances can indicate systems with issues in their abundance estimations, which serves as a warning for future investigations based on results from any large spectroscopic survey \citep[see e.g.][]{Piatti2019}.

    \item By using hierarchical clustering (HC), we selected our most metal-rich stars and further divided them into five subgroups (6 to 10). Subgroups 7 to 10 (in this order) seem to follow a pattern where all chemical abundances increase in step, suggesting a possible chemical enrichment flow (CEF), but subgroup 6 seems to have a different and/or detached CEF. Further investigation revealed that the abundances of \ion{C}{i}, \ion{Si}{ii}, \ion{Sc}{ii}, \ion{Ti}{ii}, \ion{Cr}{ii}, and \ion{Y}{ii} for the stars in subgroup 6 are affected by larger uncertainties and should be treated with care.
    
    \item By fitting isochrones, we find that most stars in all the MMR group seem to be old, with median ages of nearly 8 Gyr. Five of these stars have much lower age estimates, so we hypothesise that these systems may in fact be blue stragglers.
    
    \item By simultaneously analysing the abundance pattern of all MMR subgroups and their ages, their connection within a single CEF seems less likely. There is no clear connection in the sense that an older subgroup is also less metal-rich. On the contrary, the most metal-rich subgroup (10) is also one in which the stars are older.
    
    \item The distributions of abundances, including solar or subsolar values of [Mg/Fe] and other $\alpha$-elements, and high values of [C/H], indicate that the stars were born in regions of intense star formation, where material ejected by SN Ia, SN II, and AGB stars was present.
    
    \item The main features of our MMR sample are a chemical composition similar to thin disc stars, which indicates a complex chemical enrichment history; old ages;  $\langle Z_{\rm{max}} \rangle$ mostly above 0.3 kpc; low eccentricities with medians generally $\lesssim$ 0.2; and  metallicities that, at their guiding radius, are not consistent with the radial metallicity gradient of chemical evolution models.

    \item The above chemo-dynamic features indicate that the MMR stars in the solar neighbourhood seem to have migrated from the inner regions of the MW \citep[e.g.][and references therein]{FrancoisMatteucci1993, Grenon1987, Wilson1994, Grenon1999, SellwoodBinney2002, Andrews2017, Khoperskov2020}, which is consistent with their ages and metallicities \citep[][]{Roskar2008, Roskar2012, Roskar2013, Khoperskov2020}.
    
    \item By analysing the current guiding radii and metallicities of the stars and the distribution of eccentricities, our results indicate that the dominant dynamical process responsible for the movement of these stars from the inner Galaxy in this sample is churning, although blurring also seems to play a role. This is similar to the conclusions of  \citet{Kordopatis2015}, among others.
    
\end{enumerate}

Summarising, the MMR stars explored in this study seem to have been formed in interior regions of the Galaxy (for instance, the bulge and/or the inner disc) and, through effects of blurring and (especially) churning, have migrated to the regions closer to the solar radius. This new sample,  which has a comprehensive set of abundances in addition to the kinematic and orbital parameters,  might be ideal for future tests of chemical evolution models that include the radial motions of stars, such as those by \citet{Schonrich2009a} and \citet{Kubryk2015_01, Kubryk2015_02}.

\begin{acknowledgements}
M.~L.~L.~Dantas and R. Smiljanic acknowledge support from the National Science Centre, Poland, project 2019/34/E/ST9/00133. The authors also thank the anonymous referee for invaluable insights on the manuscript. M.~L.~L.~Dantas also thanks Miuchinha for the love and support. M.~L.~L.~Dantas thanks R. Giribaldi for the fruitful discussions. The authors thank A.~R.~da~Silva for the help with the kinematics estimates and discussions, and A.~Mints for the invaluable help with \textsc{unidam}. T.B. was funded by the project grant 2018-04857 from the Swedish Research Council. M.B. is supported through the Lise Meitner grant from the Max Planck Society, acknowledges support by the Collaborative Research centre SFB 881 (projects A5, A10), Heidelberg University, of the Deutsche Forschungsgemeinschaft (DFG, German Research Foundation), and received funding from the European Research Council (ERC) under the European Union’s Horizon 2020 research and innovation programme (Grant agreement No. 949173). This work made use of the following on-line platforms: \texttt{slack}\footnote{\url{https://slack.com/}}, \texttt{github}\footnote{\url{https://github.com/}}, and \texttt{overleaf}\footnote{\url{https://www.overleaf.com/}}. This work was made with the use of the following \textsc{python} packages (not previously mentioned): \textsc{matplotlib} \citep{Hunter2007}, \textsc{numpy} \citep{Harris2020}, \textsc{pandas} \citep{mckinney-proc-scipy-2010}. This work also benefited from \textsc{topcat} \citep{Taylor2005}. All figures were made with a qualitative palette from \url{https://colorbrewer2.org} -- credits: Cynthia Brewer, Mark Harrower, and The Pennsylvania State University. Based on data products from observations made with ESO Telescopes at the La Silla Paranal Observatory under programme ID 188.B-3002. These data products have been processed by the Cambridge Astronomy Survey Unit (CASU) at the Institute of Astronomy, University of Cambridge, and by the FLAMES/UVES reduction team at INAF/Osservatorio Astrofisico di Arcetri. These data have been obtained from the \textit{Gaia}-ESO Survey Data Archive, prepared and hosted by the Wide Field Astronomy Unit, Institute for Astronomy, University of Edinburgh, which is funded by the UK Science and Technology Facilities Council. This work was partly supported by the European Union FP7 programme through ERC grant number 320360 and by the Leverhulme Trust through grant RPG-2012-541. We acknowledge the support from INAF and Ministero dell' Istruzione, dell' Universit\`a' e della Ricerca (MIUR) in the form of the grant "Premiale VLT 2012". The results presented here benefit from discussions held during the \textit{Gaia}-ESO workshops and conferences supported by the ESF (European Science Foundation) through the GREAT Research Network Programme. This publication makes use of data products from the Wide-field Infrared Survey Explorer, which is a joint project of the University of California, Los Angeles, and the Jet Propulsion Laboratory/California Institute of Technology, funded by the National Aeronautics and Space Administration. This work has made use of data from the European Space Agency (ESA) mission {\it Gaia} (\url{https://www.cosmos.esa.int/gaia}), processed by the {\it Gaia} Data Processing and Analysis Consortium (DPAC, \url{https://www.cosmos.esa.int/web/gaia/dpac/consortium}). Funding for the DPAC has been provided by national institutions, in particular the institutions participating in the {\it Gaia} Multilateral Agreement.
\end{acknowledgements}

\bibliographystyle{aa}        
\bibliography{paper} 

\appendix

\section{Additional material}
\subsection{General properties of the Milky Way}

In this appendix we provide quick (and fairly incomplete) general parameters for the MW estimated by previous works. This is used as a quick reference to the reader with the goal of comparing the literature benchmark parameters for the MW. In Table \ref{tab:mw_properties} it is possible to see the characteristics of each substructure of the MW.

\begin{table*}
    \centering
    \caption{Below are some parameters of the different regions of the Milky Way as reference.}
    \begin{tabular}{l|r|r|l}
        Parameter       & Thin disc   & Thick disc   & References\\
        \hline
        \hline
        Thickness (kpc) & 0.3        & 0.9           & \citet[][Sec. 2.2]{McMillan2017} \\
        \hdashline
        Age (Gyr)       & $<$ 8      & $>$ 8         & \citet{Haywood2013} \\
        \hdashline
        $\rm{[Fe/H]}$ (dex)          & $\gtrsim$ $-$0.70 & $-$1.0 to $-$0.2 & \multirow{2}{*}{\citet{RB2014}} \\ 
        $\langle e \rangle$          & 0.15 & 0.3  & 
    \end{tabular}
    \label{tab:mw_properties}
    \tablefoot{LSR stands for local standard of rest.}
\end{table*}

\subsection{Additional figures and tables}

Additional plots and tables concerning the chemo-dynamic properties of the MMR subgroups are displayed in this section.

\begin{figure*}
    \centering
    \includegraphics[width=0.49\linewidth]{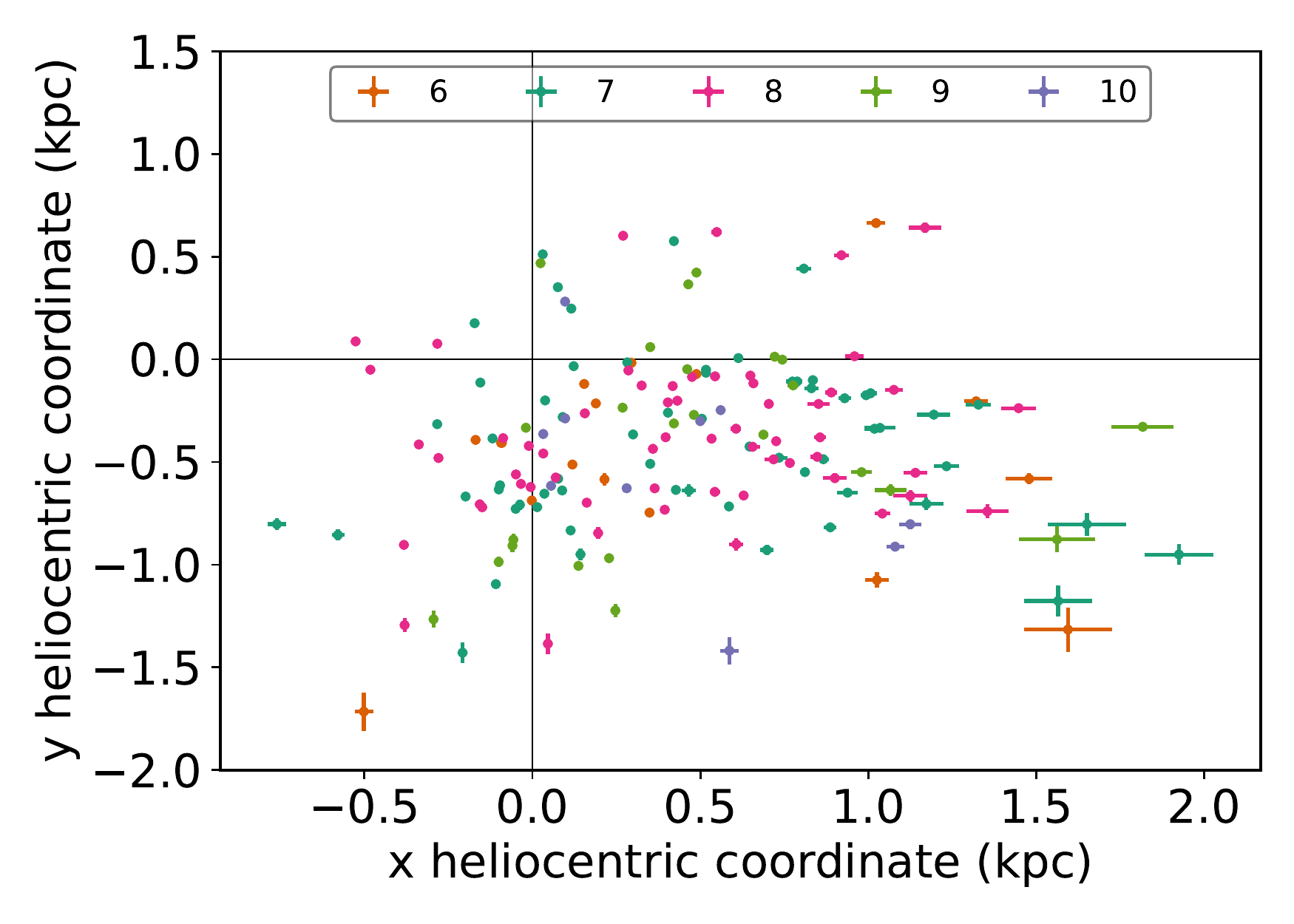}
    \includegraphics[width=0.49\linewidth]{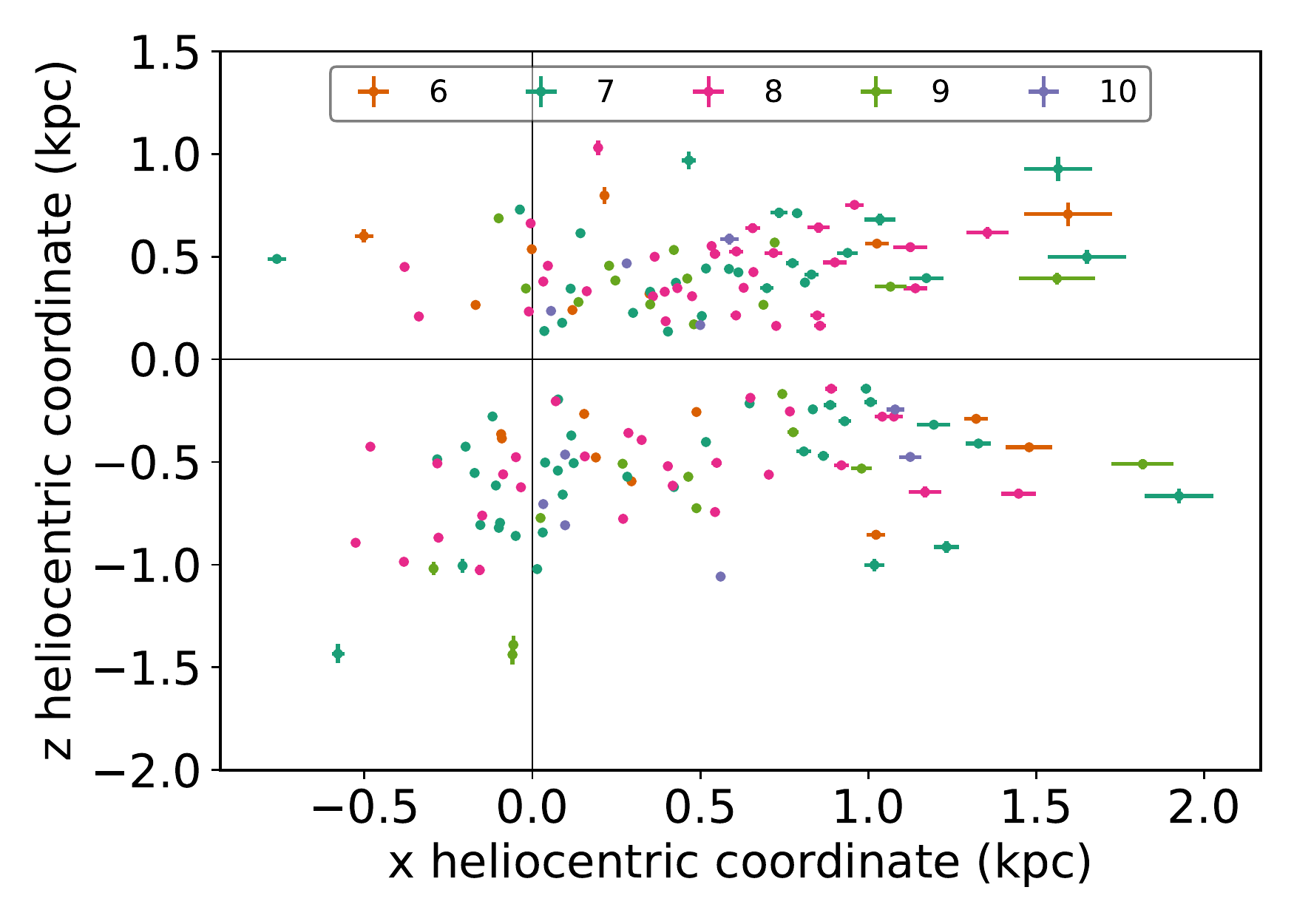}
    \caption{Heliocentric Galactic coordinates for the MMR stars. The Sun is at (x, y, z) = (0,0,0). The Galactic centre is towards negative x, according to the left-hand convention used by \textsc{galpy}. \emph{Left panel}: position of the MMR stars in $x$ and $y$ directions. \emph{Right panel}: similar to the left panel, but in the $x$ and $z$ directions.}
    \label{fig:heliocentric_distances}
\end{figure*}

Figure \ref{fig:heliocentric_distances} depicts the heliocentric rectangular projections $xy$ and $xz$ of the position of each MMR star according to their subgroup.\footnote{For more details on these parameters and others extracted from \textsc{galpy}, we refer the reader    to the reference manual: \url{https://docs.galpy.org/en/v1.7.0/reference/orbit.html}.}

Table \ref{tab:vel_disp} shows the velocity dispersion in the $z$, $r$, and $\phi$ directions for all MMR subgroups. The cylindrical velocities for all the stars were re-estimated using \textsc{astropy} since the values of $v_{\phi}$ we obtained from \textsc{galpy} seemed to be too small. We verified that $v_r$ and $v_z$ were indeed correct, but $v_{\phi}$ was much smaller than expected. We considered 8.2 kpc as the Sun's galactocentric distance \citep{McMillan2017} and 14 pc as the Sun's distance from the Galactic plane \citep{Binney1997}.

\begin{table}
    \centering
    \caption{Velocity dispersion in the $z$, $r$, and $\phi$ directions for the stars in all MMR subgroups ($\sigma_{v_z}$, $\sigma_{v_r}$, and $\sigma_{v_\phi}$).}
    \begin{tabular}{l|c|c|c|c}
        MMR       &  $\sigma_{v_z}$ & $\sigma_{v_r}$ &  $\sigma_{v_{\phi}}$ & $\overline{t}_{50\%}$\\
        subgroup  &  (km s$^{-1}$) & (km s$^{-1}$) & (km s$^{-1}$) & (Gyr)\\
        \hline
        \hline
         7 (dark green)  & 20.95 & 37.89 & 24.28 & 6.92 \\
         6 (orange)      & 23.68 & 26.15 & 27.26 & 8.91 \\
         8 (pink)        & 17.52 & 40.20 & 27.19 & 8.13 \\
         9 (light green) & 18.78 & 34.66 & 30.38 & 7.76 \\
        10 (purple)      & 23.99 & 57.89 & 25.38 & 9.12 \\
        \hline
        Total            & 19.95  & 38.29 & 27.30
    \end{tabular}
    \label{tab:vel_disp}
    \tablefoot{The order is the same as Tables \ref{tab:mmr_subgs} and \ref{tab:mmr_subgs_ages}, i.e. with decreasing values of $\langle \rm{[Fe/H]} \rangle$. We re-estimated their cylindrical velocities using \textsc{astropy} given that \textsc{galpy} provides very low estimates for ${v_{\phi}}$. To ease the comparison, we repeat their respective $\overline{t}_{50\%}$, which is also present on Table \ref{tab:mmr_subgs_ages}. To estimate the cylindrical velocities, we adopted the Sun's Galactocentric distance as 8.2 kpc \citep[see][]{McMillan2017} and $z$-height as 14 pc \citep[see][]{Binney1997}.}
\end{table}

\begin{figure*}
    \centering
    \includegraphics[width=\linewidth]{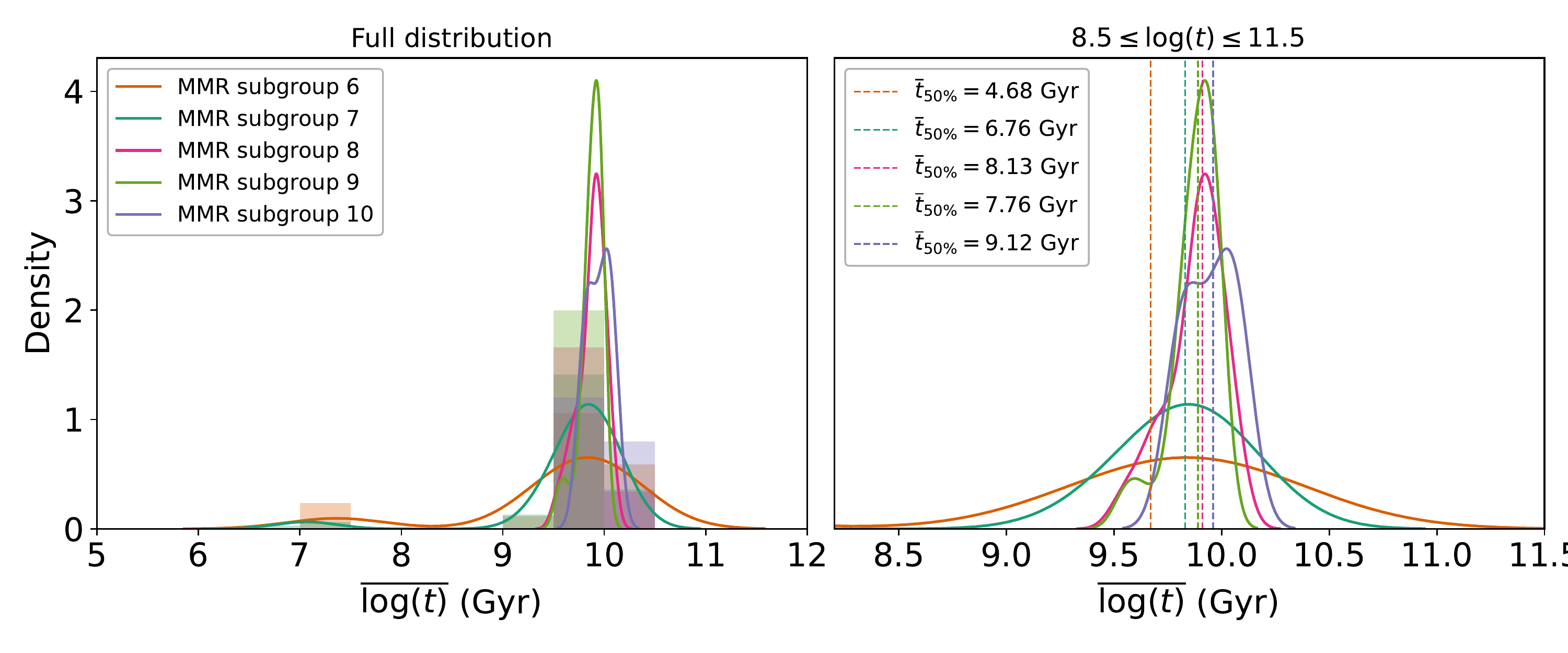}
    \caption{Distribution of ages for all MMR subgroups in terms of density. \emph{Left panel}: Entire distribution with both histograms and 1D Gaussian kernel density plots, including the outliers of younger age. \emph{Right panel}: Similar to the left panel, but without the histograms and zoomed in within 8.5$\leq \log(t) \leq$ 11.5. Additional vertical lines are added to depict the medians of each distribution (with symbol $\overline{t}_{50\%}$ for the 50${th}$ percentile of the age distribution), with the same corresponding colours for each MMR subgroup.}
    \label{fig:appendix_agedistribution}
\end{figure*}

Figure \ref{fig:appendix_agedistribution} depicts the age distributions for all the MMR subgroups. It is split into two subplots. The left one depicts the entire distribution, including the outliers with lower age, and the right one shows the same distribution zoomed in to $8.5 \leq \log(t) \leq 11.5$ and additional dashed lines with their respective medians. The depicted ages are the medians (i.e. the 50${th}$ percentile of the distribution) for the age estimations; the dashed lines are the medians of the median values of the median distributions. The median ages shown in this figure consider the outliers of younger age (see first peak on the left panel), and this is why the median ages for groups 6 and 7 (orange and dark green) are lower when compared to those in Table \ref{tab:mmr_subgs_ages}.

\begin{figure}
    \centering
    \includegraphics[width=\linewidth]{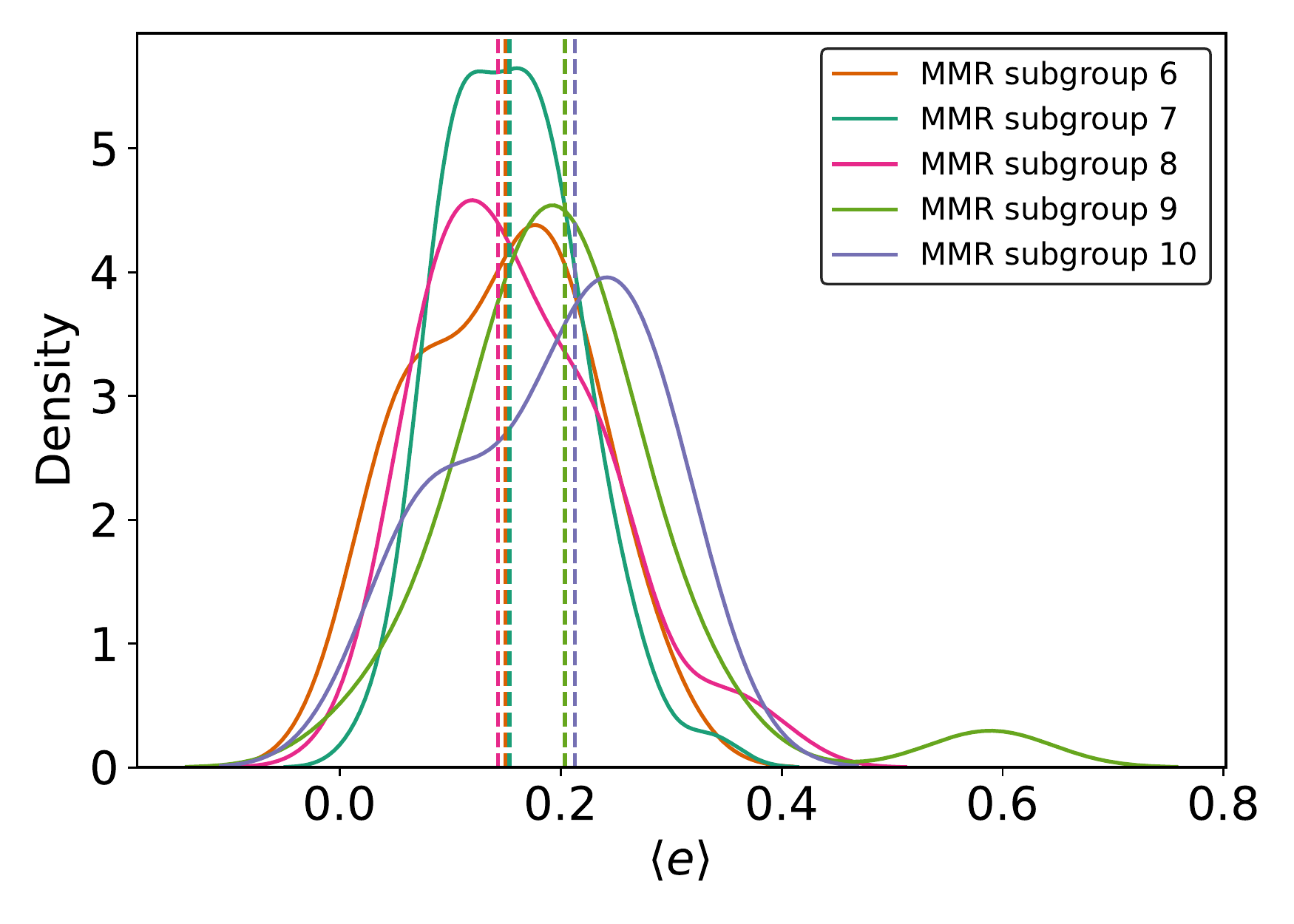}
    \caption{Median eccentricities ($\langle e \rangle$) shown as a 1D Gaussian kernel densities for all the MMR subgroups (6 to 10). Medians for each distribution displayed in dashed lines with the same colours as the corresponding distributions. Most of the stars peak at $\langle e \rangle < 0.2$, except for groups 9 and 10 that have $\langle e \rangle$ slightly higher than 2. This is in agreement with \citet[][see their Fig. 3]{Sales2009} and \citet[][see their Fig. 9]{Kordopatis2015}.}
    \label{fig:appendix_eccentricities}
\end{figure}

\begin{figure}
    \centering
    \includegraphics[width=\linewidth]{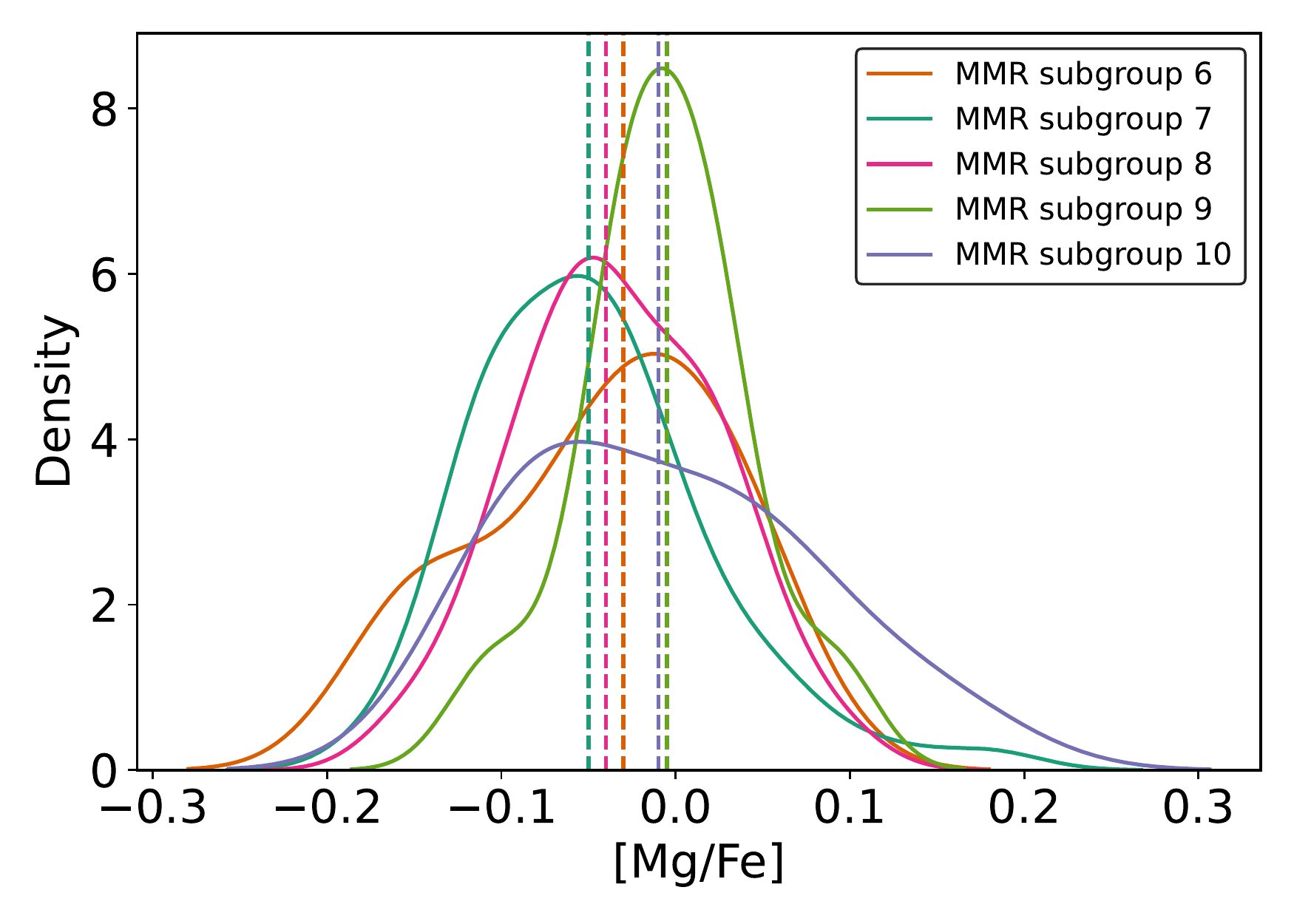}
    \caption{Median values for [Mg/Fe] shown as a 1D Gaussian kernel densities for all the MMR subgroups (6 to 10). Medians for each distribution displayed in dashed lines with the same colours as the corresponding distributions.}
    \label{fig:appendix_mgfe}
\end{figure}

Figure \ref{fig:appendix_eccentricities} displays the distributions of $\langle e \rangle$ in the shape of 1D Gaussian kernel densities. Additionally, the medians are depicted by dashed lines in the same colours as the distributions. It is possible to see that $\langle e \rangle$ peaks in eccentricities below 0.2, which is considered to be low, except for groups 9 and 10 (light green and purple) that have median eccentricities slightly larger than 0.2. Figure \ref{fig:appendix_mgfe} displays the distribution of $\langle \rm{[Mg/Fe]} \rangle$ analogously to Fig. \ref{fig:appendix_eccentricities}.

\begin{figure}
    \centering
    \includegraphics[width=\linewidth]{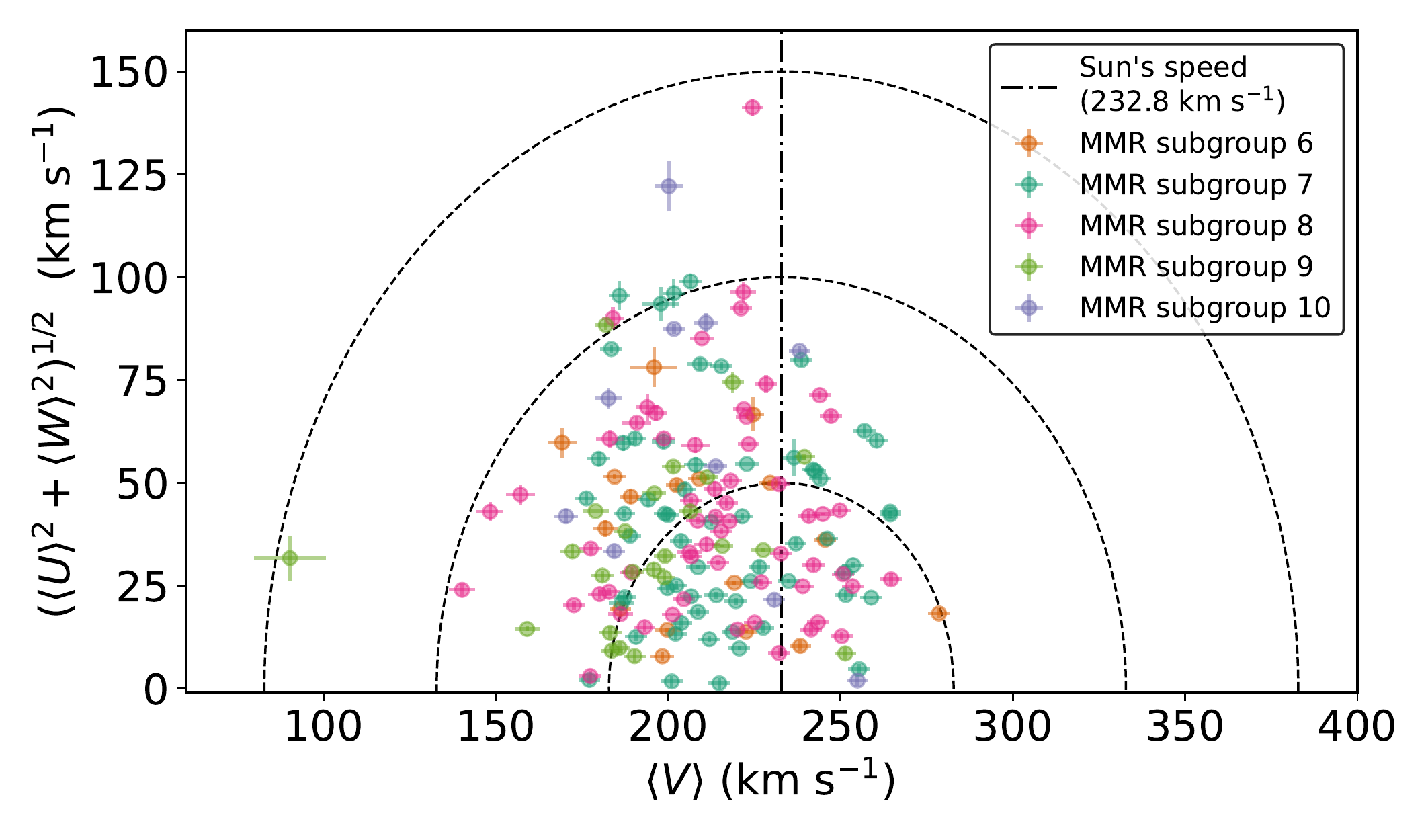}
    \caption{Toomre diagram for all the stars in the MMR group, in terms of the solar velocity, 232.8 km s$^{-1}$ \citep{McMillan2017}. All velocities are median velocities extracted from \textsc{galpy} as discussed in Sect. \ref{sec:sample}. Additionally, all velocities were corrected for the Sun's velocities \{$U_{\odot}$, $V_{\odot}$, $W_{\odot}$\}= \{8.6$\pm$0.9, 13.9$\pm$1.0, 7.1$\pm$1.0\} as estimated by \citet[][their Table 2]{McMillan2017}. The dashed curves indicate the total space velocity in concentric steps of 50 km s$^{-1}$, which is represented as $v_{\rm{tot}}=(U^2 + V^2 + W^2)^{1/2}$. The uncertainties shown in this figure have been estimated via propagation of uncertainties.}
    \label{fig:appendix_toomre}
\end{figure}

Figure \ref{fig:appendix_toomre} displays the Toomre diagram for the MMR stars. The velocities shown there correspond to the medians calculated with \textsc{galpy}. It is noticeable that all stars have prograde movement and their velocities are mostly lower than that of the Sun.

\begin{figure*}
    \centering
    \includegraphics[width=\linewidth]{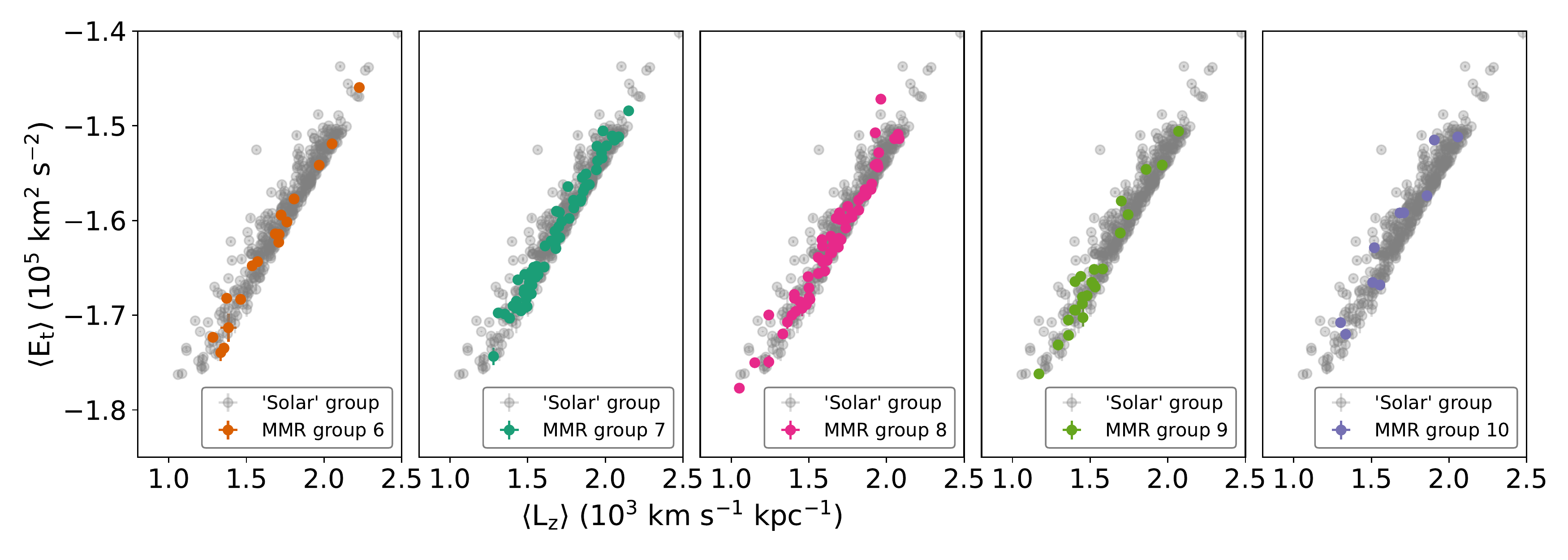}
    \caption{$\langle E_t \rangle $ \emph{\emph{vs}} $\langle L_z \rangle $ (i.e. median total binding energy \emph{\emph{vs}} median angular momentum on the $z$-axis of the MW, also known as the  Lindblad diagram) for the five MMR subgroups (in the same colours as Fig. \ref{fig:abund_lvplot}). The Solar group parameters are shown in grey for comparison. All stars seem to be in prograde movement, independently of their group classification, at several levels of binding energy.}
    \label{fig:appendix_energy_lz}
\end{figure*}

Figure \ref{fig:appendix_energy_lz} depicts the Lindblad diagram for all the stars in the MMR subgroups as well as the Solar group (in grey). All systems have a prograde trajectory (positive angular momentum, i.e. $\langle L_z \rangle$>0), but occupy different \emph{\emph{loci}} in terms of energy levels.

\end{document}